\documentclass[11pt]{article}
\usepackage{amssymb,fullpage,esvect,amsmath,graphicx,latexsym,amsthm,slashed,grffile,cite}
\usepackage[usenames,dvipsnames]{xcolor}
\usepackage{jcappub,url}
\usepackage[final]{pdfpages}
  \DeclareMathOperator{\mev}{MeV} \DeclareMathOperator{\gev}{GeV}  \DeclareMathOperator{\cm}{cm}        
       \newcommand{\cO}{{\cal O}}   
\newcommand{\ep}{\epsilon} \newcommand{\vp}{\varphi}
\newcommand{\ie}{{\it i.e.~}}  \newcommand{\eg}{{\it e.g.~}}
 \def\oL{\overline}
\newcommand{\pL}{\left(} \newcommand{\pR}{\right)} \newcommand{\bL}{\left[} \newcommand{\bR}{\right]} \newcommand{\cbL}{\left\{}  \newcommand{\mL}{\left|} \newcommand{\mR}{\right|}
\newcommand{\lm}{{\ell m}}

\DeclareMathOperator{\ks}{KS} \DeclareMathOperator{\cdf}{CDF}
\def\bea{\begin{eqnarray}}
\def\eea{\end{eqnarray}}
\newcommand{\GeV}{\text{GeV}}
\def\be{\begin{equation}}
\def\ee{\end{equation}}

\newcommand{\beq}{\begin{equation}} \newcommand{\eeq}{\end{equation}}
\newcommand{\alg}[1]{\begin{align} \begin{split} #1 \end{split}  \end{align}}

\newcommand{\Eq}[1]{Eq.~(\ref{#1})}  
\newcommand{\Sec}[1]{Sec.~\ref{#1}} \newcommand{\Secs}[2]{Secs.~\ref{#1} and \ref{#2}} 
\newcommand{\Fig}[1]{Fig.~\ref{#1}} \newcommand{\Figs}[2]{Figs.~\ref{#1} and \ref{#2}}
\newcommand{\Tab}[1]{Tab.~\ref{#1}}
\newcommand{\App}[1]{App.~\ref{#1}}

\begin{document}

\subheader{\rm FERMILAB-PUB-15-504-T, YITP-SB-15-43}
\title{Wavelet-Based Techniques for the Gamma-Ray Sky}
\author[a]{Samuel D. McDermott,}
\affiliation[a]{C. N. Yang Institute for Theoretical Physics, SUNY Stony Brook, NY, 11794}
\author[b]{Patrick J. Fox,}
\affiliation[b]{Theoretical Physics Department, Fermi National Accelerator Laboratory, IL 60510}
\author[c,d]{Ilias Cholis,}
\affiliation[c]{Department of Physics and Astronomy, The Johns Hopkins University, Baltimore, MD 21218}
\affiliation[d]{Center for Particle Astrophysics, Fermi National Accelerator Laboratory, IL 60510}
\author[e,f]{and Samuel K.~Lee}
\affiliation[e]{Broad Institute, 75 Ames Street, Cambridge, MA 02142}
\affiliation[f]{Princeton Center for Theoretical Science, Princeton University, NJ 08544}
\date{\today}

\abstract{We demonstrate how the image analysis technique of wavelet decomposition can be applied to the gamma-ray sky to separate emission on different angular scales.  New structures on scales that differ from the scales of the conventional astrophysical foreground and background uncertainties can be robustly extracted, allowing a model-independent characterization with no presumption of exact signal morphology.  As a test case, we generate mock gamma-ray data to demonstrate our ability to extract extended signals without assuming a fixed spatial template. For some point source luminosity functions, our technique also allows us to differentiate a diffuse signal in gamma-rays from dark matter annihilation and extended gamma-ray point source populations in a data-driven way.}

\maketitle
\flushbottom

\section{Introduction} \setcounter{page}{2}

Astrophysical observations are becoming increasingly precise, and we are currently at an exciting and pivotal juncture. As vast new data sets become available, we are gaining unprecedented insights into large-scale astrophysical phenomena. We are starting to better understand astronomical structures on scales ranging from local to cosmological, and some of our best-motivated theories of particle physics are finally being tested. For instance, coming generations of indirect and direct detection experiments will conceivably rule out or discover canonical classes of dark matter candidates \cite{Buckley:2013bha,Cushman:2013zza}, and the nature of both dark energy \cite{Abbott:2005bi,Levi:2013gra} and Galactic structure formation \cite{Pancino:2012aa} will begin to be revealed.

Given the systematic uncertainties that will be encountered with the impending influx of data, new analytic techniques may be required to robustly detect interesting new structures and signals. In this work, we demonstrate a technique that promises to be able to determine the presence of new signals in a variety of settings. In order to demonstrate and quantify its utility, we will work with simulated data that mimics the data acquired by the \textit{Fermi}-LAT telescope; we leave an application of this technique to actual \textit{Fermi}-LAT data to future work \cite{futurepaper}. Compared to standard template fits, our technique is less reliant on a presumptive division between foreground, background, and signal. (Henceforth, we will refer to foregrounds and backgrounds collectively as ``backgrounds.'') Instead, we decompose a given image into structures with support on different angular scales in an attempt to understand the components that are actually observed. In this way, we can reach robust conclusions about novel signals despite large uncertainties in the background models.

This separation in angular scales is accomplished at each energy bin by partitioning a given image via a wavelet transform.  Wavelets are mathematical objects that are inherently well-suited to investigations of structures at different scales \cite{1992tlw..conf.....D}. They have been successfully implemented in various image analysis settings; their use is well established in fields as varied as medical imaging, computer science, and file compression \cite{AstroImageBook}, not to mention astronomy \cite{Starck:2005fb,Schmitt:2010vf} and collider physics \cite{Rentala:2014bxa,Monk:2014uza}. Wavelets decompose an image into constituent parts in a manner that is sensitive to both the scale {\it and} the location of the image's components. This is in contrast to decompositions like the Fourier or spherical harmonic transforms, which retain information only about the {\it scales} in the original signal but are insensitive to {\it where} on an image a signal component lays. For this reason, wavelets can identify edges and other sharp structures as being qualitatively different from smooth, large-scale image components.

The twin expectations that the gamma-ray sky is composed of structures with support on different angular scales and that structures on these different scales have different astrophysical origin are relatively robust. The expected diffuse gamma-ray emission from the Milky Way is the result of interactions of high-energy cosmic rays with background microwave, infrared, and starlight photons as well as Galactic gas that has been mapped in other frequency ranges. In addition, point sources in the Galaxy, extragalactic star-forming galaxies, active galactic nuclei, and misidentified cosmic rays contribute to the total gamma-ray background emission. As we will show, an analysis of the sky that is sensitive to different scales will separate photons associated to these processes from the more broadly varying signal that would be produced by a new source or a collection of sources arranged smoothly in the Galaxy. Possible sources of smooth, extended emission include dark matter annihilation, a collection of unresolved point sources, or a transient outflow of cosmic rays in the recent past. Furthermore, these different sources of diffuse emission may themselves be characterized by a concentration of power on their own unique scales. If this is true, it would open the possibility of characterizing a putative excess discovered in the data in a model-independent way (\eg without fitting to an assumed morphology).

We will apply the scale-sensitivity of the wavelet decomposition to identify in a data-driven way the angular scales that support interesting signals. Our analysis relies on no specific morphological features or particular choice of background models. Instead, we use the breadth of systematic uncertainties on the backgrounds to provide a threshold for establishing the significance of new signal components.

The remainder of the paper is organized as follows. In \Sec{sec:Wavelets} we introduce the wavelet transform and discuss the technical details of our implementation of the wavelet mechanism. In \Sec{sec:Data} we give an overview of the data sets of interest.  We describe our set of background models and we outline an algorithm for extracting interesting new signals that depends on the spread of these background maps. In \Sec{sec:Sensitivity} we proceed to provide a discussion of expected sensitivity to various gamma-ray signals. In \Sec{sec:Conclusion}, we describe some future applications of our methodology, and we conclude.

\section{Wavelets}
\label{sec:Wavelets}

The main technical tool on which we base our analysis is the wavelet transform. By way of introduction, in \Sec{wavelet-technical} we discuss the continuous and discrete one dimensional wavelet transforms, with emphasis on the discrete \emph{\'a trous} transform; a more complete introduction and overview of wavelets can be found in \cite{AstroImageBook,keinertbook}.  In \Sec{IUWTS} we introduce a generalization of the {\it \'a trous} transform for use on the sphere, which will be the basis of the algorithm we introduce in \Sec{sec:Data}. In \Sec{uncertainties} we discuss the how uncertainties propagate through the wavelet transform. The reader who is strictly interested in the technicalities necessary for implementing the algorithm introduced in \Sec{sec:Data} need only focus on \Sec{IUWTS}.

\subsection{Features of the Wavelet Transform}
\label{wavelet-technical}

A wavelet is a square integrable function $\psi$ whose translations and rescalings also provide a basis of square integrable functions. An example of a continuous wavelet is the ``Mexican Hat" wavelet where, up to a normalization factor, $\psi_s(x) \propto (1-x^2/s^2) e^{-x^2/s^2}$. A signal $f$ can be integrated against this wavelet at positions $p$ and scale sizes $s$ to provide a set of wavelet coefficients,
\be
w(p,s) = \frac{1}{\sqrt{s}}\int_{-\infty}^{\infty}dx f(x)\psi^*\left(\frac{x-p}{s}\right)~.
\label{eq:CWT}
\ee
The original signal can be reconstructed from the wavelet coefficients by the inverse transform,
\be
f(x) = \frac{1}{N_\psi} \int_0^\infty\frac{ds}{s^2}\int_{-\infty}^{\infty}db\frac{1}{\sqrt{s}} w(p,s)\psi\left(\frac{x-p}{s}\right)~,
\label{eq:CWTinv}
\ee
where the normalization is related to the integral of the square of the Fourier transform of the wavelet:
\be
N_\psi = \int_0^{\infty} \frac{d\nu}{\nu}|\hat{\psi}(\nu)|^2.
\ee
We define the Fourier transform as $\hat{\psi}(\nu)=\int_{-\infty}^{\infty} dt \psi(t) e^{-2\pi i \nu t}$.  Notice that in order for the inverse transform to be defined the wavelet function must have a mean of zero \ie\ $\hat{\psi}(0)=\int dt\, \psi(t)=0$.  There are an infinity of square integrable functions of mean zero that one could choose for the wavelet function, depending on the desired characteristics of the decomposition.  In what follows we will use a simple wavelet function based on a smooth third-order polynomial which we define below.

As mentioned earlier, an essential feature of the wavelet transform is that the wavelet coefficients provide information about both the central location and the spatial extent of the structures in the original signal. In the astrophysical imaging context, wavelets effectively identify the scale of different emission components, even if they are superimposed on one another. Identifying the structures that are present only on the lowest wavelet levels allows us to tag noise and small scale structures and remove these from our image. By removing filaments and point sources in this way, only smooth large-scale emission components should remain. Using a wavelet-based technique, we can robustly improve the signal to background ratio in residual images and offer a model-independent characterization of new emission components that does not rely on the existence of a presumed signal template.

Just as Fourier transforms have a continuous and discrete version there is a discrete wavelet transform where the integrals in \Eq{eq:CWTinv} can be replaced with summations.  The summation over position can be made finite if the system has appropriate boundary conditions.  It is most useful to replace the integral over scales with a dyadic sum over scales which are $2^{n}$ ($n\ge 0$) larger than the resolution of the data, length scale $L$.  To make the summation over scales finite we restrict the sum, $n< N$, where $N$ is bounded by the requirement that $2^N L$ should be smaller than the largest length scale in the data.  It is necessary to introduce one additional function, the so-called \emph{scaling function} $\phi$, whose purpose is to capture the average of the signal on length scales $2^N L$ or larger.  Thus, the discrete wavelet decomposition of $f$ is:
\be
f=\sum_k c_{N,k}\phi_{N,k} + \sum_{n=1}^{N-1} \sum_{k} w_{n,k} \psi_{n,k}~,
\label{eq:DWTdecomposition}
\ee
where $\psi_{n,k} = 2^{\frac{n}{2}} \psi(2^n x-k)$, and similarly for $\phi$. In the continuous version the transform and its inverse are usually determined by the same function.  For the discrete wavelet transform it is often beneficial to loosen this requirement and introduce dual functions for $\phi$ and $\psi$, $\tilde{\phi}$ and $\tilde{\psi}$.  These satisfy a bi-orthgonality condition such that $\int \tilde{\psi}_{ni}\psi_{mj}=\delta_{mn}\delta_{ij}$, and similarly for $(\phi,\,\tilde{\phi})$.  However, in practical applications it is not always necessary to know the form of the dual functions.  The discrete wavelet transform has the benefit that it can be recast as a filtering procedure, allowing fast algorithms on computers.  

The scaling and wavelet functions satisfy recursion relations,
\be
\phi(x)=\sum_k \sqrt{2} h_k \phi(2x-k) \quad \psi(x)=\sum_k \sqrt{2} g_k \psi(2x-k)~.
\ee
From these it follows that the $c$, $w$ in \Eq{eq:DWTdecomposition} can also be determined recursively,
\be
c_{N,j} = \sum_k h_{k-2j}c_{N-1,k} \quad  w_{N,j} = \sum_k g_{k-2j}c_{N-1,k}~.
\label{eq:filtereqns}
\ee
These recursion relations are seeded by assuming that the original data is a discrete sampling of the signal using $\phi$, \ie the data points are taken to be $c_{0,k}$.

From a practical point of view \Eq{eq:filtereqns} is sufficient to define an algorithm to carry out the wavelet decomposition of any signal.  The discrete transform we will utilize is a generalization of the so-called \emph{\'a trous} algorithm (also known as the stationary or undecimated wavelet transform) \cite{AstroImageBook}, in which the smoothing of the image is implemented by multiplying each pixel by a matrix with increasingly sparse support at different wavelet levels. The {\it \'a trous} algorithm is ``redundant,'' in the sense that the {\it \'a trous} decomposition contains more data than the original signal.  Canonical wavelet transforms like the Haar wavelet reduce the dimensionality of the target space at each wavelet level; this ``decimation'' can allow compression of data and images.  The redundancy in the \emph{\'a trous} algorithm, on the other hand, is useful for preserving position information between wavelet levels, since every level of the decomposition has the same number of basis elements. This is useful for analysis of images such as the astrophysical data we consider presently. Because every level of the undecimated transform has the same number of basis elements but contains modes with different support, such transforms can alternately be viewed as a set of generalized bandpass filters.

\subsection{Isotropic Undecimated Wavelet Transforms on the Sphere}
\label{IUWTS}

The discussion of one-dimensional wavelet transforms is straightforward to generalize to higher dimensions: one may simply carry out the convolutions of \Eq{eq:filtereqns} along orthogonal Cartesian directions simultaneously.  However, given the symmetries of the gamma-ray sky we prefer to carry out the transform in a basis of spherical harmonics.
For this purpose, we utilize the isotropic undecimated wavelet transform on the sphere (IUWTS) \cite{Starck:2005fb}. The IUWTS decomposes scales in a manner similar to the {\it \'a trous} wavelet transform discussed above, but the smoothing step in the IUWTS decomposition is accomplished by removing the highest modes of the spherical harmonic decomposition rather than by respacing the elements of the initial smoothing matrix. The smoothing and wavelet functions naturally live in spherical harmonic space, and they only carry dependence on the $\ell$ mode: they are isotropic in the sense that they carry no azimuthal angular information. This ensures that the different levels of the wavelet transform inherit azimuthal dependence solely from the initial map.

\begin{figure}[t]
\begin{center}
\includegraphics[width=.6\textwidth]{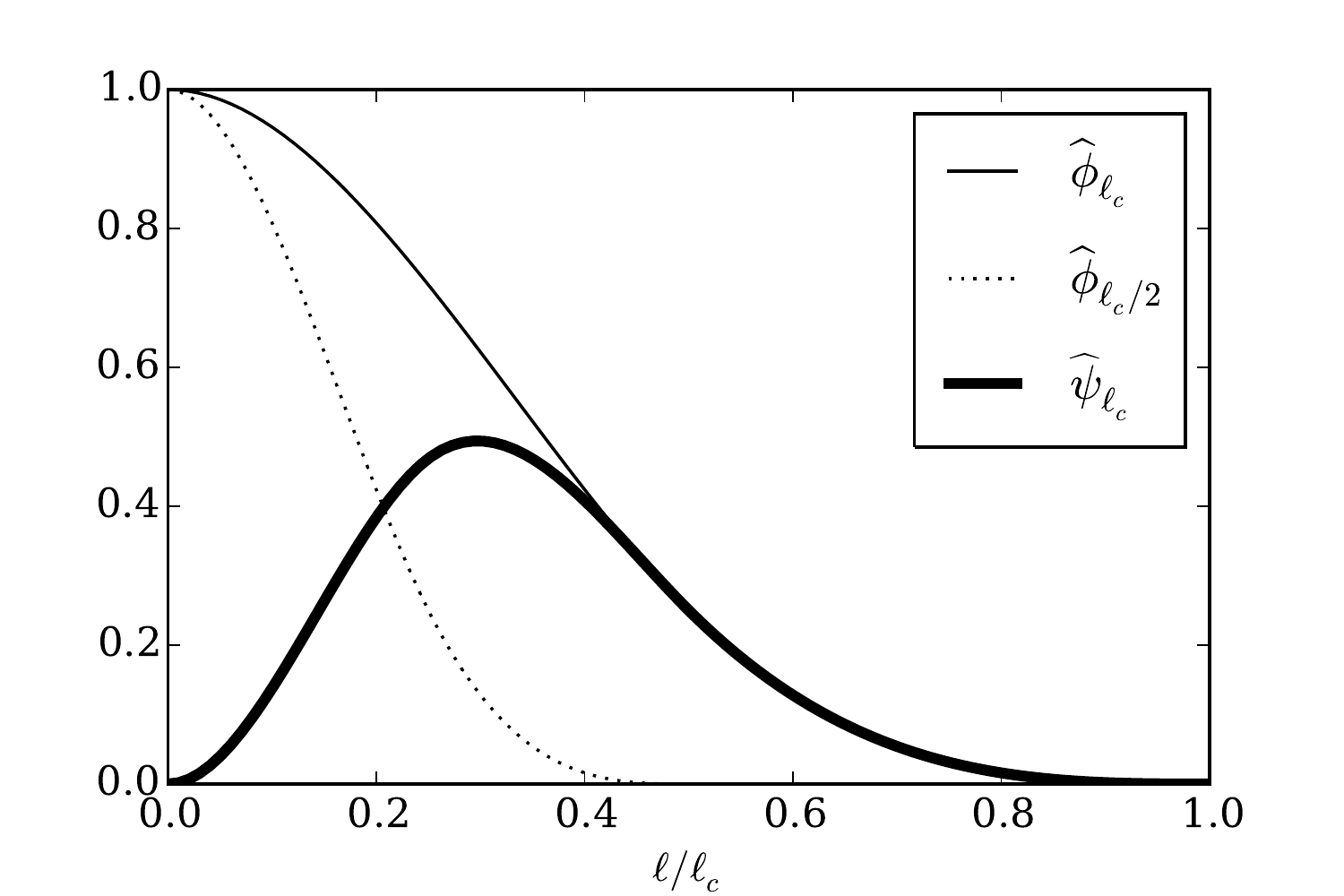}
\caption{The spherical harmonic transforms of the scaling function $\hat \phi$ and wavelet function $\hat \psi$ as a function of mode number $\ell$ in units of the maximum mode number $\ell_c$. The matrices $\hat H$ and $\hat G$ from \Eq{H and G def} are defined at discrete values of $\hat \phi$ and $\hat \psi$.}
\label{wav-scal}
\end{center}
\end{figure}

We begin by expanding the scaling and wavelet functions in spherical harmonics,
\be
\phi_{\ell_c}(\theta,\phi)=\sum_{\ell=0}^{\ell=\ell_c} \hat{\phi}_{\ell_c}(\ell)Y_{\ell0}(\theta,\phi).
\ee
(Here and in the following, we use a hat to denote the spherical harmonic transform.) Following \cite{Starck:2005fb}, we take our scaling function $\hat{\phi}_{\ell_c}$ to be related to the cardinal $B$-spline of order 3. The $B$-splines are defined recursively, using:
\beq
B_k(x) = \int_{-\infty}^{\infty} dz B_0(z-x) B_{k-1}(z)\quad \mathrm{and}\quad B_0(x)=\Theta(1/2+x)\Theta(1/2-x),
\eeq
where $\Theta(x)$ is the Heaviside step function. The closed form of $B_3$ is:
\be
B_3 ( x) = \frac1{12} \pL \mL x+2 \mR^3 -4 \mL x+1 \mR^3 +6\mL x \mR^3  -4 \mL x -1 \mR^3 + \mL x - 2 \mR^3\pR  \Theta \pL 4-x^2 \pR.
\ee
The scaling and wavelet functions, defined in spherical harmonic space, are:
\be
\hat \phi_{\ell_c}  (\ell) = \frac32 \, B_3 \! \left(\frac{2\ell}{\ell_c} \right) \qquad \hat  \psi_{\ell_c}  (\ell) = \hat  \phi_{\ell_c}  (\ell) - \hat  \phi_{\ell_c/2}  (\ell).
\ee
The function $\hat \phi$ is normalized to 1 for $\ell=0$ and approaches zero very quickly as $\ell \to \ell_c$. We plot $\hat \phi$ and $\hat \psi$ in \Fig{wav-scal}.

As discussed earlier, it is possible to determine the wavelet coefficients by recursively applying high/low pass filters to the signal.  Starting from the original image $M$, the wavelet decomposition begins by smoothing $M$ at some very high wavenumber. This gives a first smoothed map $c_0(M;\theta, \vp)$ with spherical harmonics $\hat c_0(M;\ell, m)$. In what follows, we will typically drop the label $M$ from $c_j$ and $w_j$, although all smoothed maps and all wavelet levels are to be understood as being obtained from some original map or image $M$. Increasing levels of the IUWTS are recursively defined by:
\beq \label{wav-def}
\hat c_{j+1}(\ell,m) = \hat c_j(\ell,m) \hat H_j(\ell),\qquad \hat w_{j+1}(\ell,m)=\hat c_j(\ell,m) \hat G_j(\ell) = \hat c_j(\ell,m) - \hat c_{j+1}(\ell,m),
\eeq
up to a maximum level $j_{\rm max}$. The maximum level is determined by the number of modes $\ell_{\rm max}$ retained in the first smoothed image $c_0$, with $j_{\rm max}=1+\log_2\ell_{\rm max}$.  The filters, $\hat H$ and $\hat G$, are $\ell \times (2\ell+1)$ matrices defined in terms of the scaling and wavelet functions as:
\beq \label{H and G def}
\hat H_j(\ell,m) \equiv \frac{\hat \phi_{\ell_c/2^{j+1}}(\ell)}{\hat \phi_{\ell_c/2^j} (\ell)}  \Theta\pL \frac{\ell_c}{2^{j+1}} - \ell \pR , \qquad\qquad \hat G_j(\ell,m) \equiv  1 - \hat H_j(\ell,m) .
\eeq
Again, we point out that the IUWTS inherits azimuthal information solely from the image.

\begin{table}[t]
\begin{center}
\begin{tabular}{|c||c|c|c|c|c|c|} \hline
$j$ & $w_1$ & $w_2$ & $w_3$ & $w_4$ & $w_5$ & $w_6$ \\ \hline
$\theta$ & $[0.7^\circ,1.4^\circ]$ & $[1.4^\circ,2.8^\circ]$ & $[2.8^\circ,5.6^\circ]$ & $[5.6^\circ,11.3^\circ]$ & $[11.3^\circ,22.5^\circ]$ & $[22.5^\circ,45^\circ]$ \\ \hline
\end{tabular}
\end{center}
\caption{Angular scales that dominate the wavelet levels $w_j$ for $\ell_{\rm max}=512$.}
\label{Wavelet scales}
\end{table}%

For any $j$, the ``wavelet level'' $w_j$ depicts objects with support in the modes which are being removed as one goes to the image $c_j$. The most smoothed map depicts the (isotropic) average of the original image. Each wavelet level leading up to $c_{j_{\rm max}}$ depicts structures of a fixed range of sizes, increasing dyadically with $j$. Like the \'a trous wavelet, the IUWTS offers lossless reconstruction of an initial image $c_0$ by adding together all wavelet levels $w_j$ with the monopole $c_{j_{\rm max}}$:
\beq
c_0(\theta,\vp) = \sum_{j=1}^{j_{\rm max}} w_j(\theta,\vp) + c_{j_{\rm max}}(\theta,\vp).
\eeq
Because each level of the wavelet decomposition has the same number of pixels, the IUWTS contains redundant information. The maps $c_j$ (obtained by repeated application of $\hat H$) are smoothed by removing high-$\ell$ modes from $M$. The levels $w_j$ (obtained by repeated application of $\hat G$, or equivalently by differences of adjacent values of $c_j$) allow resolution of small-scale structures by subtracting smoothed images from less smooth images.

The IUWTS is inherently suited to the study of isotropic emission. For instance, other wavelet transforms like the ridgelet or curvelet transforms can break the isotropy of the transform and pick out a preferred axis~\cite{Starck:2005fb,Chan:2015jiv}. In contrast, the IUWTS provides a mild bias in favor of isotropic and homogeneous structures, since infinitely many modes are required to reconstruct arbitrarily hard edges; by dropping wavelet levels, we automatically smooth the edges of the underlying structure. However, we consider this bias to be outweighed by the model-independent nature of the IUWTS, since, in the absence of some principled reason to break isotropy by picking an axis, assuming isotropy is a safe starting point.

We emphasize that this procedure is different from Gaussian smoothing (or other averaging) over small angular scales. When a wavelet level $w_j$ is removed, structures whose angular support is confined to the angular scales between the levels $j-1$ and $j$ are removed entirely, and do not contribute any power to the smoothed image $c_j$. In this way, the IUWTS is formally similar to implementing a series of bandpass filters. However, the smoothness of the scaling function $\hat \phi$ ensures that no structures are ever caught ``between levels,'' as would happen if $\hat \phi$ were a simple step function. The IUWTS in some sense provides a generalized ``soft filter,'' where the filter is not sharply sensitive to the cutoffs of each level.

\subsection{Uncertainties}
\label{uncertainties}

There are two sources of uncertainty inherent to this method. First, each observed pixel carries with it a Poisson uncertainty from the statistics of counting. Second, the variety of possible diffuse maps introduces an independent uncertainty due to the different choices of cosmic ray injection and diffusion parameters. We identify these as statistical and systematic errors, respectively.

We can largely remove Poisson noise, which has support primarily at the single-pixel scale, simply by ignoring the wavelet level $w_1$. However, because the IUWTS takes global information and is inherently nonlocal, these Poisson errors do not entirely disappear from the final image. We describe the propagation of statistical errors through the wavelet transform in more detail in \App{stat unc}.

After reducing counting fluctuations, we would additionally like to reduce the systematic uncertainty inherited from the wide variety of background templates. Crucially, the background templates we utilize differ most strongly on small angular scales (where observations are difficult), but roughly agree at the largest angular scales. Thus, in order to reduce systematic uncertainties in a minimally biased fashion, we would like to establish a data-driven method for selecting the wavelet levels of interest that preferentially weights the large angular scales where background understanding is relatively strong. We describe such a method in the following section.

\section{Data Analysis and Methodology}
\label{sec:Data}

As mentioned above, we will model our analysis on mock data inspired by the \textit{Fermi} Gamma-Ray Space Telescope \cite{1999APh....11..277G}, although the wavelet techniques described here are by no means restricted to gamma-ray data. The richness of the gamma-ray sky and the thorny nature of its systematic uncertainties allow a nice demonstration of the advantages of these wavelet-based techniques.

The presence of structures bright in GeV photons is expected based on complementary observations at lower frequencies. The tracers for the GeV structures are seen in different bands depending on the source and target population. These maps are used for calculations of cosmic-ray propagation and also to produce the expected gamma-ray background for a given set of assumptions on the properties of the interstellar medium. For instance, inelastic collisions of hadronic cosmic rays with the Galactic gas can produce neutral pions that decay to pairs of photons. The map of photons from these $\pi^0$-initiated photons is the integral along the line of sight of the $\pi^{0}$ component gamma-ray emissivity, which in turn is proportional to the product of the gas density and the hadronic cosmic ray density. If instead leptonic cosmic rays interact with the gas, gamma-rays from bremsstrahlung emission will be produced. The leptonic cosmic rays can also up-scatter background starlight, infrared or CMB radiation into gamma-rays.

The \textit{Fermi}-LAT has observed the gamma-ray sky with unprecedented precision since 2008. The sky is indeed rich with gamma-ray photons, providing extensive opportunities for learning about high-energy astrophysics from our own galaxy. The majority of these high-energy photons originate, as expected, in the disk of the Milky Way. A hard component of GeV emission that is well-separated from the Galactic disk, known colloquially as the ``\textit{Fermi} Bubbles'' \cite{Dobler:2009xz, Su:2010qj,Fermi-LAT:2014sfa} has been observed at large angles, becoming the dominant source at high energy. Claims of an even brighter component of GeV gamma-rays originating closer to the plane of the disk \cite{Goodenough:2009gk,Hooper:2010mq,Hooper:2011ti,Abazajian:2012pn,Hooper:2013rwa,Gordon:2013vta,Abazajian:2014fta,Daylan:2014rsa,Calore:2014xka,TheFermi-LAT:2015kwa} but extending outwards over more than $10^\circ$ \cite{Calore:2014xka}, will be discussed in more detail below.

In order to have some sensitivity to the different spectral shapes expected for the various background components and putative signals we divide the photons into four energy bins: $[0.5,1.0]\  \GeV$, $[1.0,2.2]\  \GeV$, $[2.2,4.9]\  \GeV$, and $[4.9,10.8]\  \GeV$.  Within each energy bin we construct a flux map of the sky and obtain wavelet coefficients for levels $2-7$, corresponding to angular scales of $\sim 2^\circ - 90^\circ$ (see Table~\ref{Wavelet scales}).  We determine these coefficients by applying the IUWTS on the unmasked sky, and then ``zooming in on'' a region of interest defined by some opening angle around the Galactic center.  Throughout this paper we make use of {\tt HEALPix}\footnote{Available at \url{http://healpix.sourceforge.net}.} with {\tt NSide=128}, corresponding to 196,608 pixels of $\sim 0.5$ square degrees.

\subsection{Diffuse Templates}
\label{sec:Diffuse Templates}

In the context of a known background, we would have excellent sensitivity to dark matter, even without a wavelet-based approach. Using wavelets to remove Poisson noise would only improve the sensitivity: statistical fluctuations have support mostly on the single-pixel scale, and coherent oscillations of the background are much more rare, so the uncertainty would exist predominantly on the first wavelet level. At larger angular scales, the statistical uncertainty is smaller, and any new large scale emission component should be easy to identify.

Unfortunately, the gamma-ray sky is not understood at this level of detail. The systematic uncertainties on the background components dominate the Poisson sampling errors. Identifying a new component of emission is therefore an exercise in determining the correct way to account for the systematic uncertainty on the background templates. We use a wide class of diffuse templates to parameterize the spread of backgrounds and to help determine whether or not an image contains a large-scale excess. To realistically account for the admissible variations in the Galactic diffuse emission, we use 19 simulated sets of maps of the Galactic gamma-ray sky that were produced using {\tt GALPROP} \cite{Strong:2007nh}. These are models A-D, F-R, W and GXI from \cite{Calore:2014xka}, which probe the main uncertainties in the Galactic diffuse emission relevant to the work at hand (for more details, see discussion in Sec.~3 and App.~A of \cite{Calore:2014xka} as well as \cite{Ackermann:2012pya}). 

\begin{figure}[tbp]
\begin{center}
\includegraphics[width=.3\textwidth]{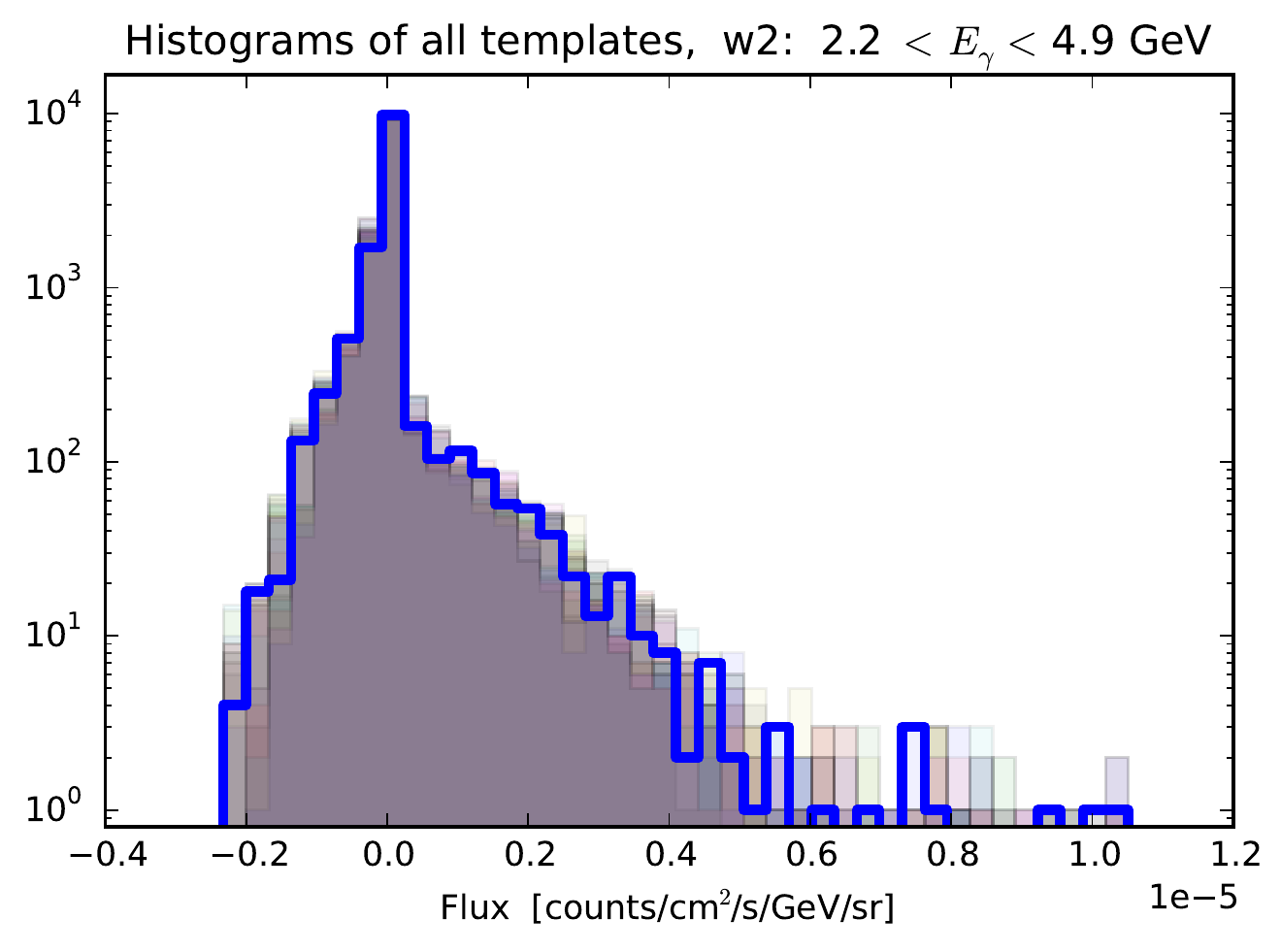}~~~
\includegraphics[width=.3\textwidth]{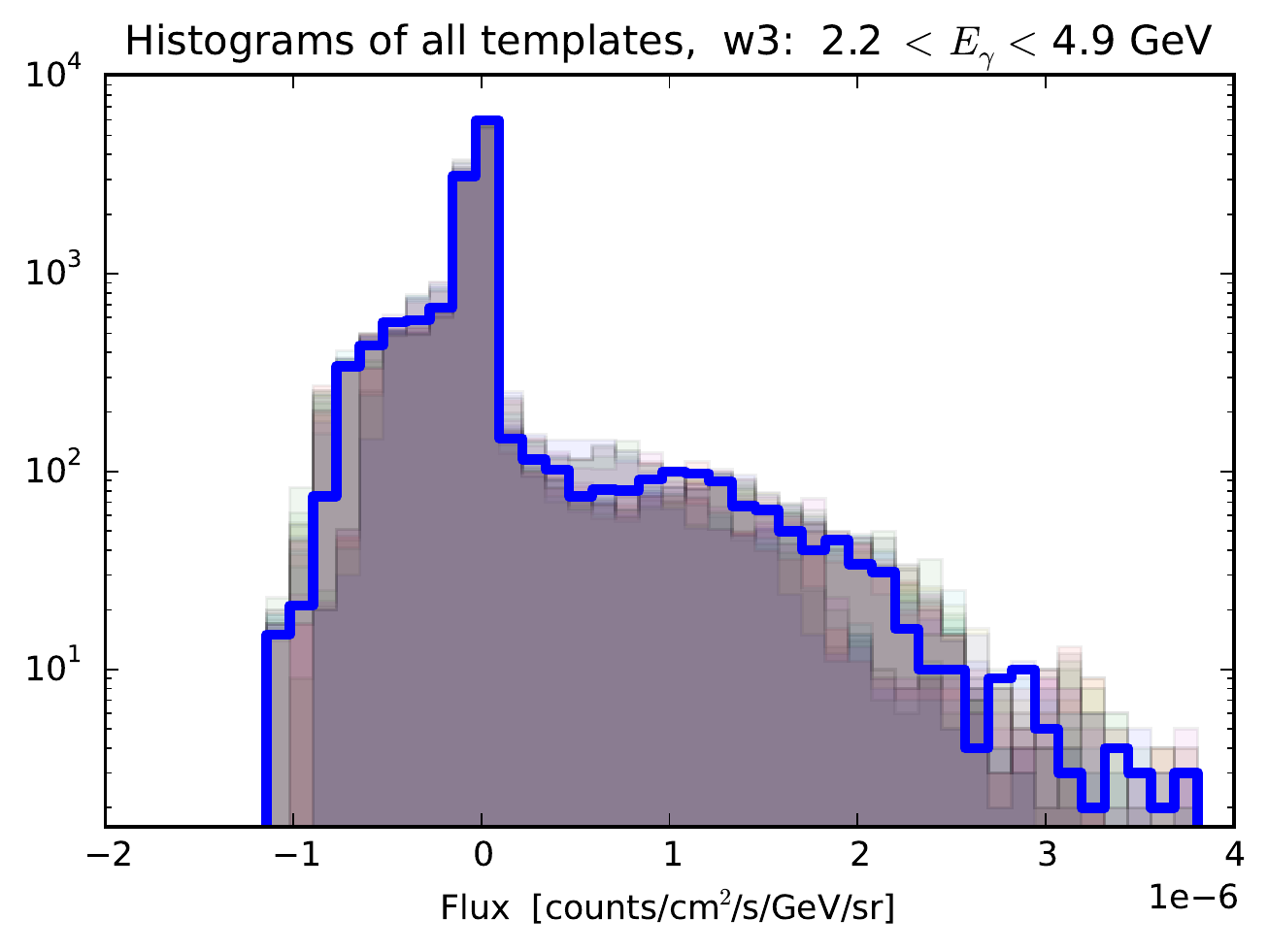}~~~
\includegraphics[width=.3\textwidth]{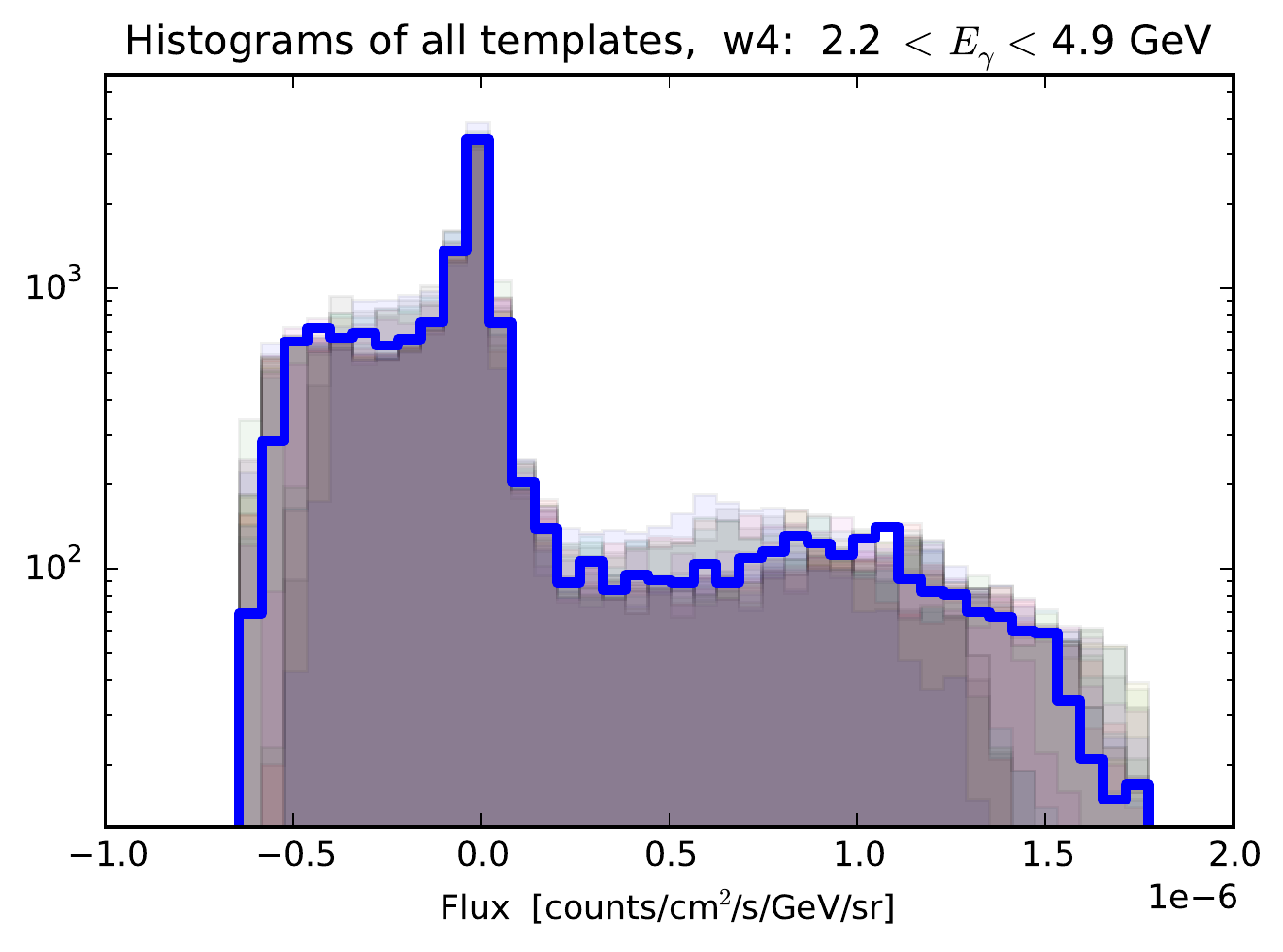}
\includegraphics[width=.3\textwidth]{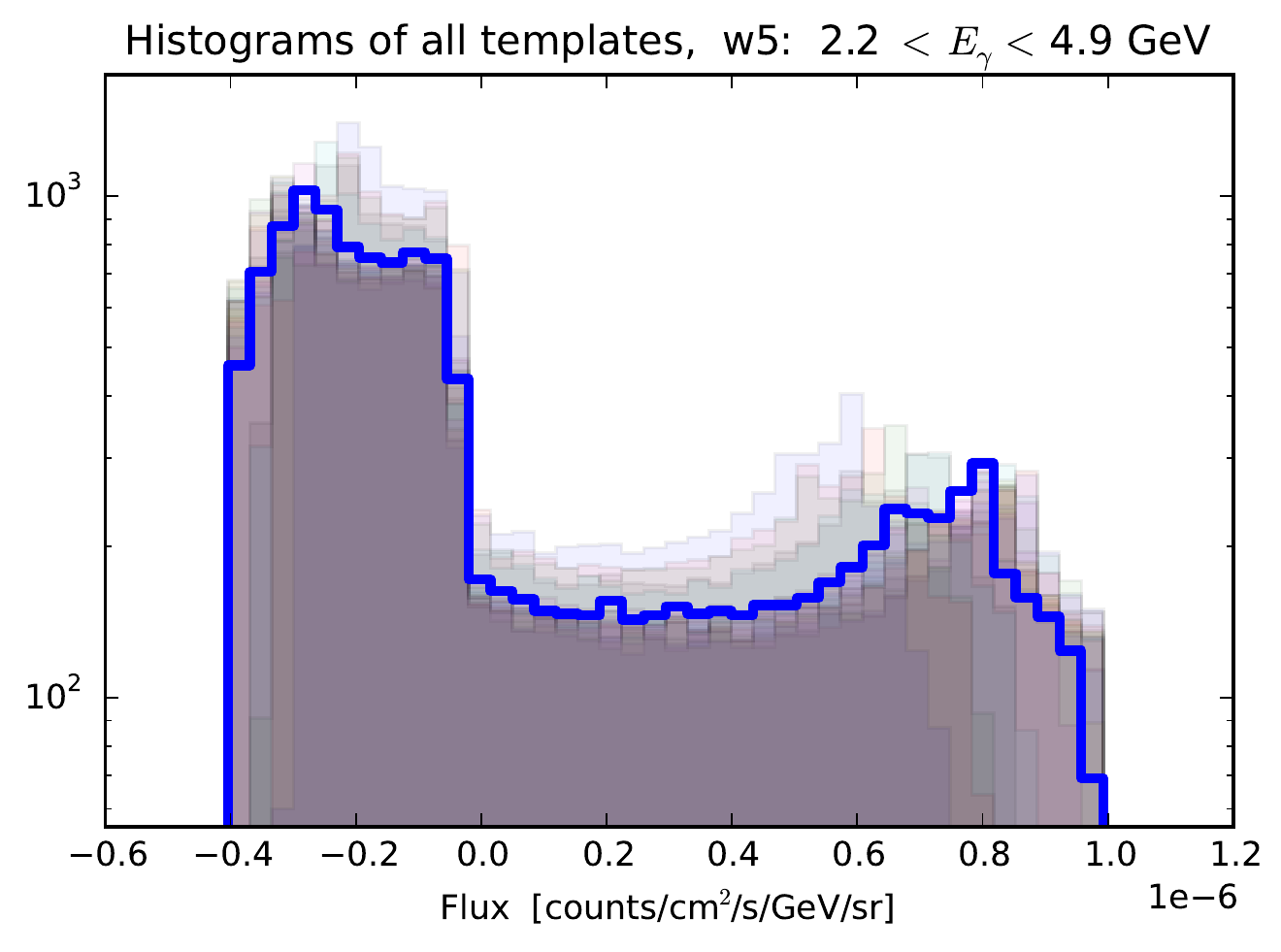}~~~
\includegraphics[width=.3\textwidth]{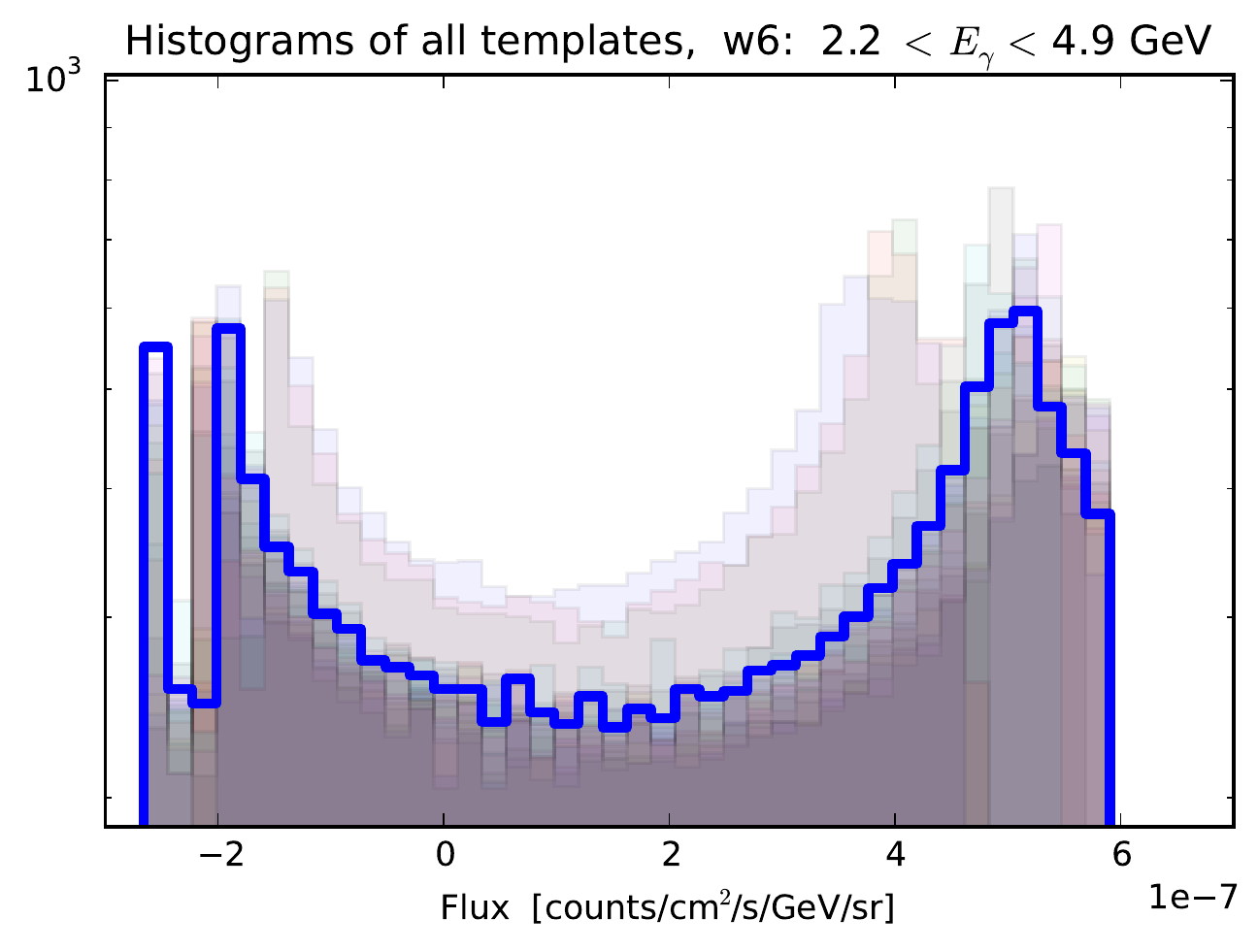}~~~
\includegraphics[width=.3\textwidth]{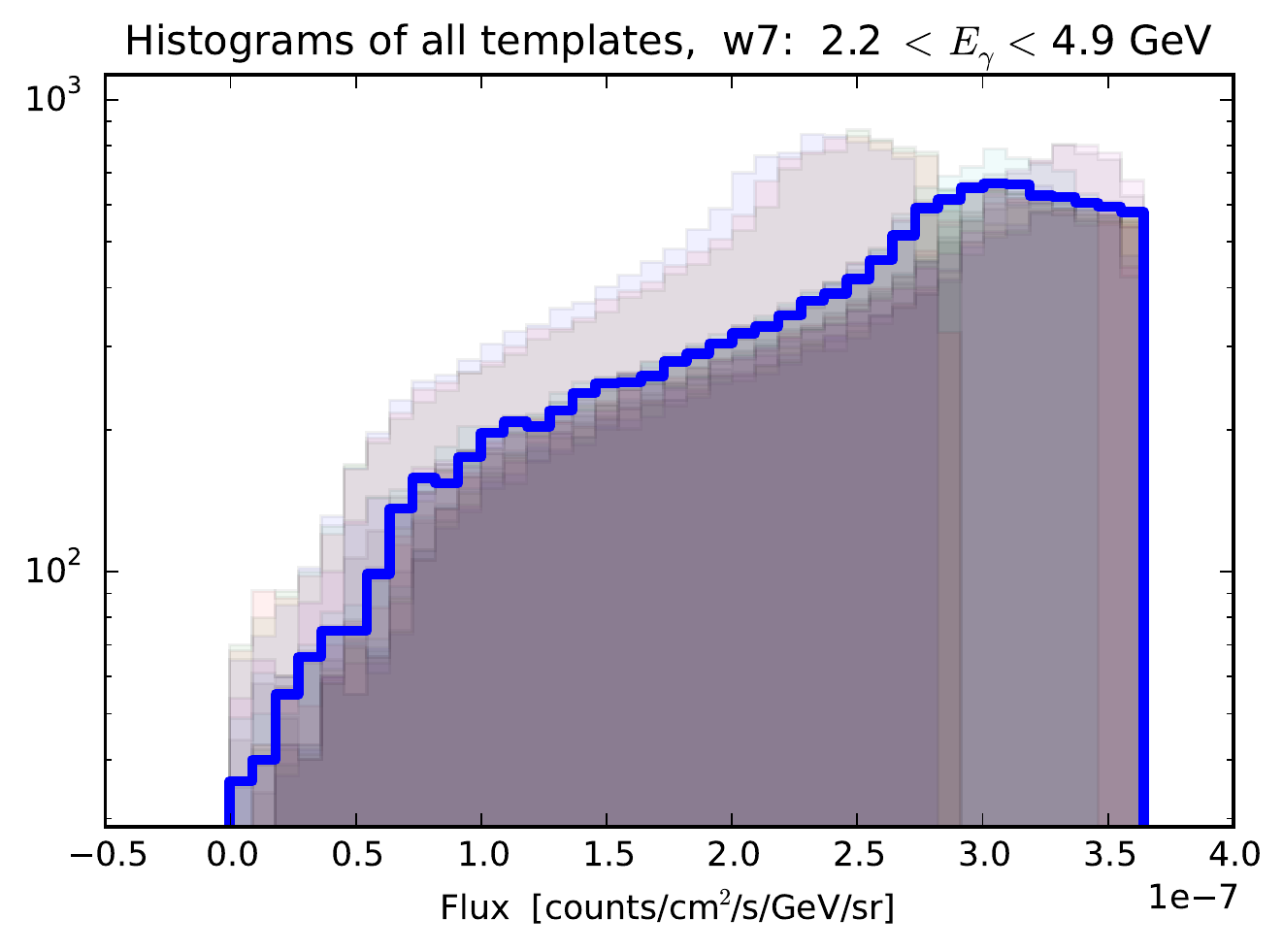}
\caption{Histograms of flux per pixel within $30^\circ$ of the Galactic center for all background templates (light shaded) and for the average background CDF, $\langle\cdf\rangle$, (thick blue) for wavelet levels 2 through 7, binning photons with $2.2\leq E_\gamma\leq4.9\gev$. See \Secs{sec:Diffuse Templates}{sec:Cleaning Maps} for more details.}
\label{bgd histograms}
\end{center}
\end{figure}

In \Fig{bgd histograms} we indicate the level of variation between the background templates. We consider the third energy bin, $[2.2,4.9]\  \GeV$, and look at the distribution of wavelet coefficients from the flux maps for wavelet levels $2-7$ in a region of interest inside $30^\circ$ from the Galactic center.  The $x$-axis in each frame is the flux in units of counts/cm${}^2$\!/s/GeV/sr, and the $y$-axis is the number of pixels in the template with a given flux. Each individual background template is shown as a translucent shaded histogram, while the average cumulative distribution function (described in more detail in \Sec{sec:Cleaning Maps}) is shown as the thick blue line. From this figure, it is evident that there are substantial systematic uncertainties encompassed by the set of background models we have retained. It is also clear that the variations are more pronounced at lower angular scales; at the scales with support on $w_6$ and $w_7$, the maps have mostly bifurcated to two basic classes (determined the spin temperature parameter \cite{Ackermann:2012pya}) without much variation within each class.

\subsection{Cleaning Maps to Find Structures}
\label{sec:Cleaning Maps}

We are interested in whether or not a given data set includes any ``excess'' emission components. As described in detail above, this is difficult because we do not know the exact interstellar conditions underlying the gamma-ray emission in the sky, providing a large spread of possible models for the Galactic diffuse backgrounds. Having qualitatively characterized our uncertainties about the astrophysical diffuse emission, we need a quantitative measure of a significant excess beyond these uncertainties. In this section, we describe a thresholding procedure to determine wavelet levels of interest, where the threshold is obtained from the spread of the background models. As described here, we will use the Kolmogorov--Smirnov test to determine this threshold at each wavelet level.

The Kolmogorov--Smirnov (KS) test is a nonparametric method for comparing two data sets. The KS test provides a quantitative measure of the probability that two data realizations $M_a$ and $M_b$ come from the same underlying distribution \cite{KS-orig1,KS-orig2}. The test statistic is defined as:
\beq
\ks\pL M_a,M_b\pR = \sup_x \mL \cdf\bL M_a\!\pL x\pR \bR - \cdf\bL M_b\!\pL x\pR \bR \mR,
\eeq
which is the maximum distance between the empirical cumulative distribution functions ($\cdf$) of the two data sets.  Here $\sup$ is the supremum, $x$ is the single independent variable that parametrizes the two data sets $M_a$ and $M_b$, and $\cdf\bL M(x) \bR$ is the cumulative distribution function of the data $M$ as a function of $x$. In our case, $x$ represents the photon flux, while $M_a$ and $M_b$ each represent a data map, an expected map generated from templates, or the wavelet levels thereof.  We carry out the KS test for each energy bin and wavelet level.

$\ks(M_a,M_b)$ is always bounded between 0 and 1. A value of $\ks\pL M_a,M_b\pR$ that is close to 0 indicates that there is a high probability that the models $M_a$ and $M_b$ are realizations of the same underlying distribution. Increasing values of the KS test represent {\it monotonically decreasing probabilities} that two data sets are samples from the same underlying distribution.  To carry out the comparison of two data sets all that is needed are the CDFs constructed from the full data sets. For each wavelet level $j$ and energy bin $E$, the background model CDF that we will compare against is an average over all 19 background models.  This background CDF, denoted $\langle\cdf\rangle$, is formed by concatenation of every pixel of every set of diffuse templates we use. $\langle\cdf\rangle$ is an appropriately normalized CDF which is equal to the average of the CDFs of all individual background templates.

With this average background CDF, we endeavor to build up a clean residual map from the wavelet decomposition of a given signal map $S$. We must first determine the wavelet levels that differ significantly from the expectation. Our procedure is as follows:
\begin{enumerate}
\item we compute the KS test values for the signal versus $\langle\cdf\rangle$, $\ks(S,\langle\cdf\rangle;j,E)$, for each $j$ and $E$.
\item at each wavelet level, we also compute a set of 19 KS test values for the diffuse templates themselves\footnote{Specifically, these are obtained from Poisson realizations of each of the 19 background templates $B_i(E)$ that went into forming $\langle\cdf\rangle$.}, $\ks(B_i,\langle\cdf\rangle;j,E)$, where $1\leq i\leq19$ labels the diffuse templates described in the preceding section.  In combination with the KS value determined in the previous step, we thus obtain a set of twenty Kolmogorov--Smirnov test values at each wavelet level and energy bin which we denote as $\oL\ks(S;j,E)$.
\item for a given signal $S(E)$, we define:
\beq \label{wjgreater}
W_j(S;E) = \cbL \begin{array}{ll} w_j(S;E) & {\rm~if~}\ks(S,\langle\cdf\rangle;j,E){\rm~exceeds~a~significance~threshold} \\ 0 & {\rm~else}. \end{array} \right. 
\eeq
\end{enumerate}
By definition, wavelet levels for which $W_j \neq 0$ are levels that differ significantly from the average background, so we refer to these as ``significant wavelet levels'' in the following. The threshold we adopt in the following is that $W_j$ is significant if it is in the top $40\%$ of $\oL\ks(S;j,E)$. With these steps, we have described a ``hard-thresholding'' procedure~\cite{AstroImageBook} for isolating the wavelet levels on which a deviation from the expectation is most likely to be an interesting signal, as suggested by the spread of background modeling uncertainties.

With these significant wavelet levels in hand, we define a cleaned map that contains only the significant wavelet levels,
\beq \label{clean}
C(S;E) = \sum_{j=2}^{j_{\rm max}} W_j(S;E).
\eeq
Note that in \Eq{clean} we omit $w_1$ in all cases, since this consists almost exclusively of noise. We also create a background analogue of $C$ using the average of the background templates, $\oL B(j,E) = \sum_{i=1}^{19} B_i(j,E)/19$ where we add only the wavelet levels $j$ at which $W_j(S;E) \neq 0$. Formally, this cleaned background map is defined as:
\beq
C_B(\oL B,S;E)\equiv \sum_{j=2}^{j_{\rm max}} w_j(\oL B;E) \times \Theta\bL W_j(S;E)\bR.
\eeq
This provides a background whose uncertainties capture the expected range of possibilities, as considered by \cite{Calore:2014xka}.  Note that we include only those wavelet levels on which the variations between background models are less drastic than the disagreement of the data with the average background. We finally are interested in the cleaned residual map,
\beq \label{resid}
\Delta C(\oL B,S;E) = C(S;E) - C_B(\oL B,S;E).
\eeq
The residual map from \Eq{resid} is the primary output of our analysis. The cleaned residual map $\Delta C(\oL B,S;E)$ reveals structures that are unlikely to arise from a misunderstanding of the Galactic diffuse emission. This method is data- and background-modeling-driven, and relies on no assumptions about the signal morphology or energy spectrum. We obtain a clean residual image using only disagreements with the background expectation on angular scales which have relatively robust background expectations.

\section{Expected Sensitivity to Large-Scale Structures in the Data}
\label{sec:Sensitivity}

The mock data we generate includes galactic diffuse emission selected from a template listed in \Sec{sec:Diffuse Templates}, the {\it Fermi} 3FGL associated point source catalog \cite{Acero:2015hja}, and an isotropic emission component. In addition, our mock data includes simulated signals that allow us to simulate the sensitivity of the procedure described in \Sec{sec:Cleaning Maps}. We simulate a variety of different signals, allowing us to investigate the possibility of using wavelets to discriminate not only between background and signal, but even between different signal components. In \Tab{tab:SimulationsTab}, we summarize the details of all the signals we have simulated.

As a demonstration of our method, we establish expected sensitivities for three possible classes of gamma-ray signals: smooth extended emission from around the Galactic center, a new Galactic center population of gamma-ray point sources, and the high-latitude \textit{Fermi} Bubbles. We treat these sky regions in order.

\begin{table*}[h]
\begin{center}
    \begin{tabular}{c| cccccc}
         \hline
Sim. Name & Galactic Diffuse & Bubbles & $M \cdot \sigma v$ & $\alpha$  & cutoff\\
            \hline \hline
            DM35 & ${}^{\rm S}$O${}^{\rm Z}$10${}^{\rm R}$30${}^{\rm T}\!$150${}^{\rm C}$5 & \checkmark & 30 & $\alpha_E=1.5$ & $E_c =7$ GeV  \\
            DM35 (dim) & ${}^{\rm S}$O${}^{\rm Z}$10${}^{\rm R}$30${}^{\rm T}\!$150${}^{\rm C}$5 & \checkmark & 9 & $\alpha_E=1.5$ & $E_c =7$ GeV  \\ \hline
            PS1.2 & ${}^{\rm S}$S${}^{\rm Z}$6${}^{\rm R}$20${}^{\rm T}\!$150${}^{\rm C}$5 & - & - & $\alpha_L=1.2$ & $L_c =1.0\times10^{34}$ erg/s  \\
            PS1.5 & ${}^{\rm S}$S${}^{\rm Z}$6${}^{\rm R}$20${}^{\rm T}\!$150${}^{\rm C}$5 & - & - & $\alpha_L=1.5$ & $L_c =1.0\times10^{34}$ erg/s   \\
            PS2.0 & ${}^{\rm S}$S${}^{\rm Z}$6${}^{\rm R}$20${}^{\rm T}\!$150${}^{\rm C}$5 & - & - & $\alpha_L=2.0$ & $L_c =1.0\times10^{34}$ erg/s   \\ \hline
            Bubbles & ${}^{\rm S}$O${}^{\rm Z}$10${}^{\rm R}$30${}^{\rm T}\!$150${}^{\rm C}$5 & \checkmark & -  & - & -  \\
            \hline \hline
            \end{tabular}
\end{center}
    \caption{Summary of the simulated data. $M \cdot \sigma v$ is in units of $10^{-26}\cm^3\!$/s, and $M \sim \cO(30)$ is the approximate photon multiplicity per dark matter annihilation for 35 GeV dark matter.  Each of $M$, $\alpha_E$, $\alpha_L$, $E_c$, and $L_c$ is defined in \App{app:Simul}. For the point sources, we assume a millisecond pulsar-like energy spectrum and take each simulation to have a total luminosity of 1.3-1.8$\times 10^{37}$ erg/s.}
    \label{tab:SimulationsTab}
\end{table*}

\subsection{The Inner Galaxy}

The excess signal towards the Galactic center and Inner Galaxy has been increasingly well characterized \cite{Daylan:2014rsa,Goodenough:2009gk,Hooper:2010mq,Hooper:2011ti,Abazajian:2012pn,Hooper:2013rwa,Gordon:2013vta,Abazajian:2014fta,Calore:2014xka,TheFermi-LAT:2015kwa}. Intriguingly, the morphology of the excess appears to match the shape expected from dark matter annihilation, falling off smoothly as a function of the distance from the center of the galaxy approximately as the square of an NFW profile. This has led to a flurry of activity in both the particle physics \cite{Alves:2014yha,Logan:2010nw,Boehm:2014hva,Modak:2013jya,Huang:2013apa,Okada:2013bna,Hagiwara:2013qya,Buckley:2013sca,Anchordoqui:2013pta,Buckley:2011mm,Boucenna:2011hy,Marshall:2011mm,Zhu:2011dz,Buckley:2010ve,Berlin:2014tja,Izaguirre:2014vva,Agrawal:2014una,Cerdeno:2014cda,Ipek:2014gua,Ghosh:2014pwa,Boehm:2014bia,Ko:2014gha,Abdullah:2014lla,Martin:2014sxa,Berlin:2014pya,McDermott:2014rqa,Huang:2014cla,Balazs:2014jla,Wang:2014elb,Cheung:2014lqa,Detmold:2014qqa,Arina:2014yna,Han:2014nba,Cline:2014dwa,Basak:2014sza,Hardy:2014dea,Guo:2014gra,Yu:2014pra,Cahill-Rowley:2014ora,Banik:2014eda,Bell:2014xta,Okada:2014usa,Frank:2014bma,Hooper:2012cw} and astrophysics \cite{Carlson:2012qc,Cholis:2012fr,Carlson:2014nra,Fields:2014pia,Cholis:2014fja,Silverwood:2014yza, Agrawal:2014oha,Geringer-Sameth:2014yza,Brooks:2014qya,Cholis:2014lta,Cholis:2014noa,Cirelli:2014lwa,Zhou:2014lva,Bringmann:2014lpa,Portillo:2014ena,Petrovic:2014uda,Cholis:2015dea,Carlson:2014cwa,Yoast-Hull:2014cra,Bernal:2014mmt,Bramante:2014zca,Drlica-Wagner:2014yca,Cholis:2013ena,Hooper:2014ysa,Geringer-Sameth:2014qqa,Tavakoli:2013zva, Gaggero:2015nsa, Lee:2014mza, 
Lee:2015fea,Bartels:2015aea,Brandt:2015ula,Carlson:2015ona, Linden:2015qha, Dutta:2015ysa,deBoer:2015kta} communities, and competing interpretations of the source of the excess abound. While the existence of a signal uncorrelated with well-established astrophysical gamma-ray sources has been uncontroversially demonstrated, at this point the detailed nature of the signal, not to mention its interpretation, remain subjects of intense debate.

One of the most exciting applications of our wavelet-based algorithm will be to uncover data-driven evidence and a model-independent characterization of this excess \cite{futurepaper}. Because the systematic uncertainties dominate the statistical errors~\cite{Calore:2014xka}, our technique may have much to offer. Here, we will focus on mock data on sky regions inside an opening angle $\psi\leq30^\circ$; in this analysis, we will not mask the Galactic disk.

\begin{figure}[tbp]
\begin{center}
\includegraphics[width=.3\textwidth]{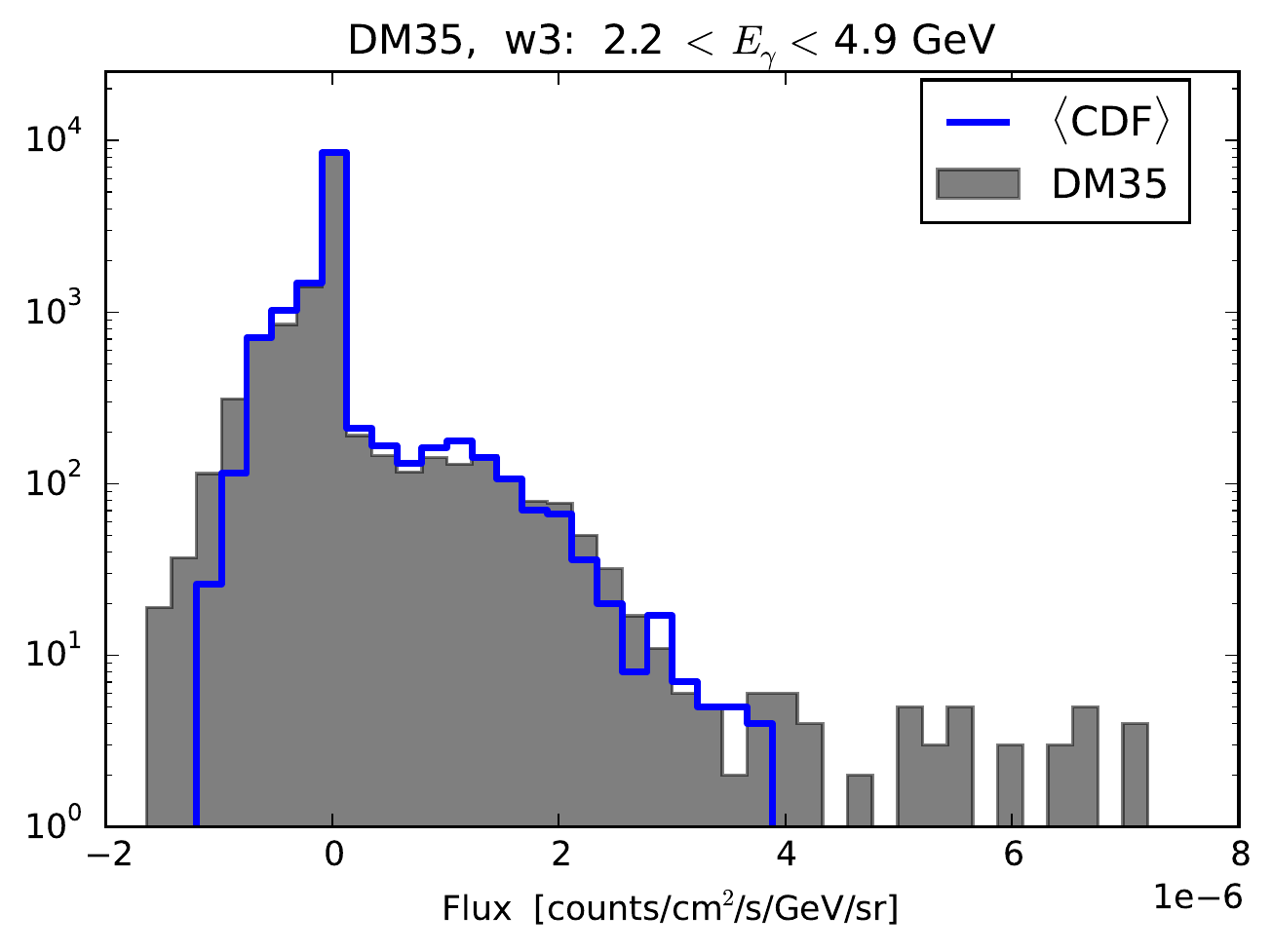}~~~
\includegraphics[width=.3\textwidth]{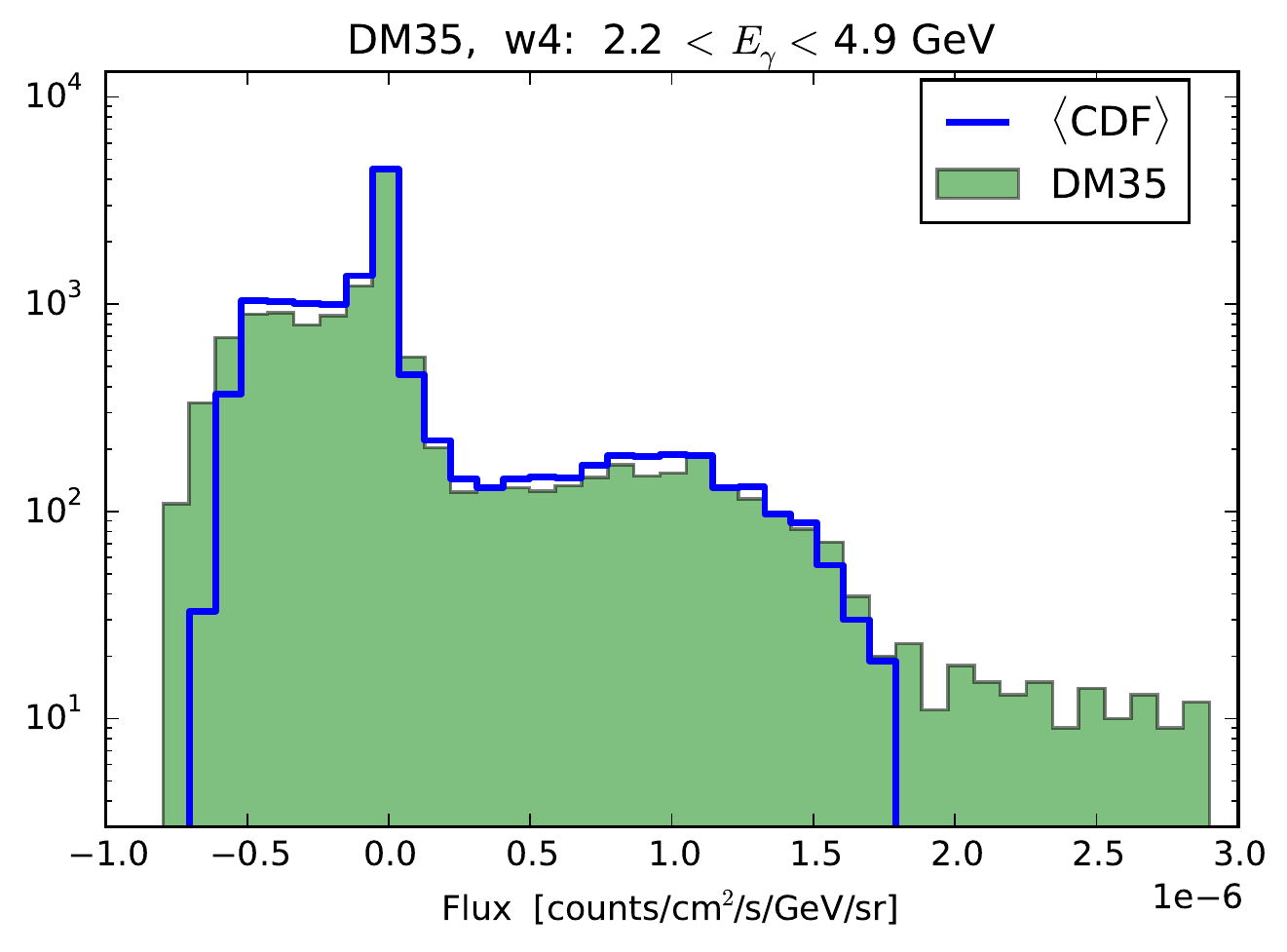}~~~
\includegraphics[width=.3\textwidth]{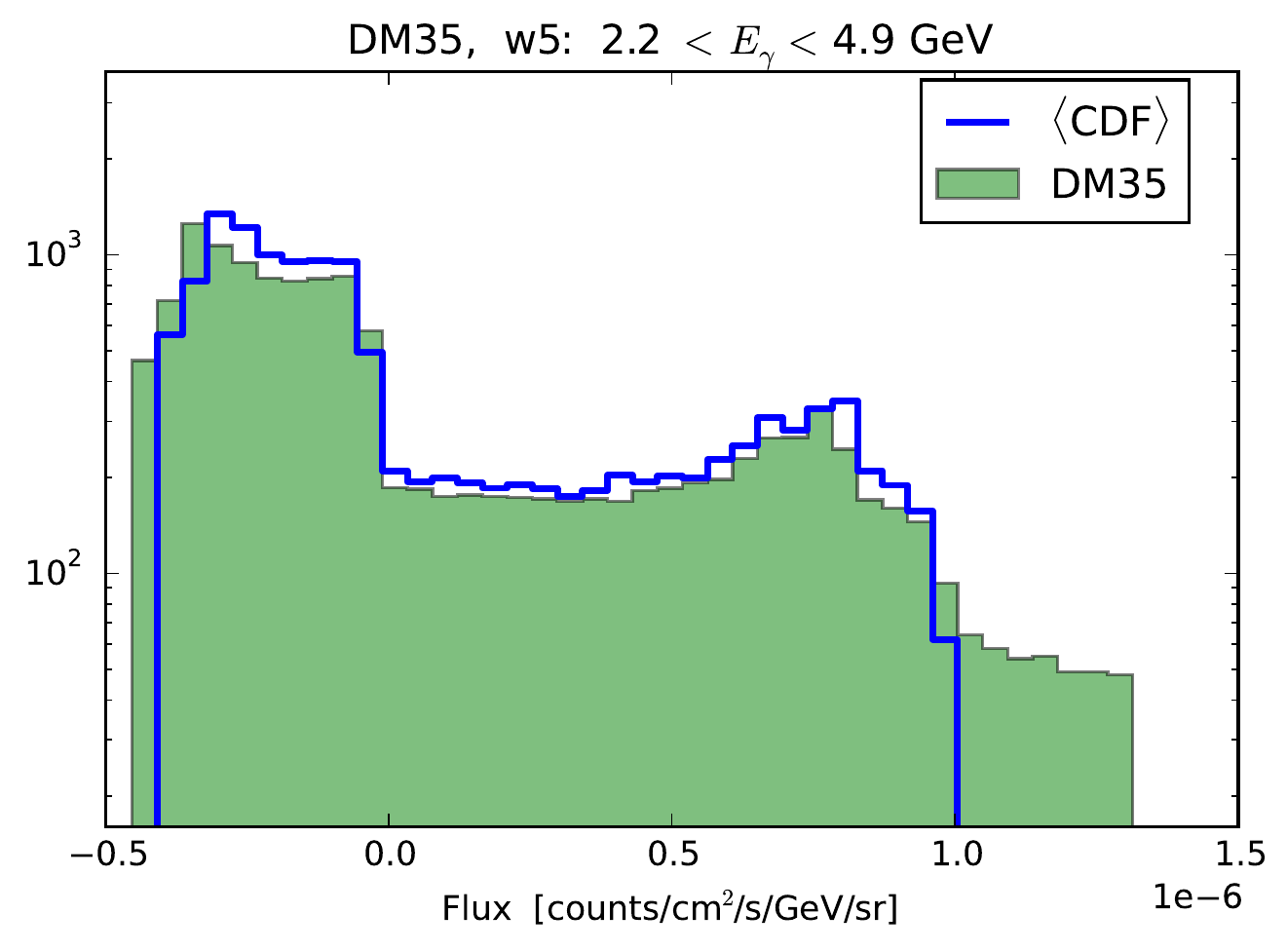}
\includegraphics[width=.3\textwidth]{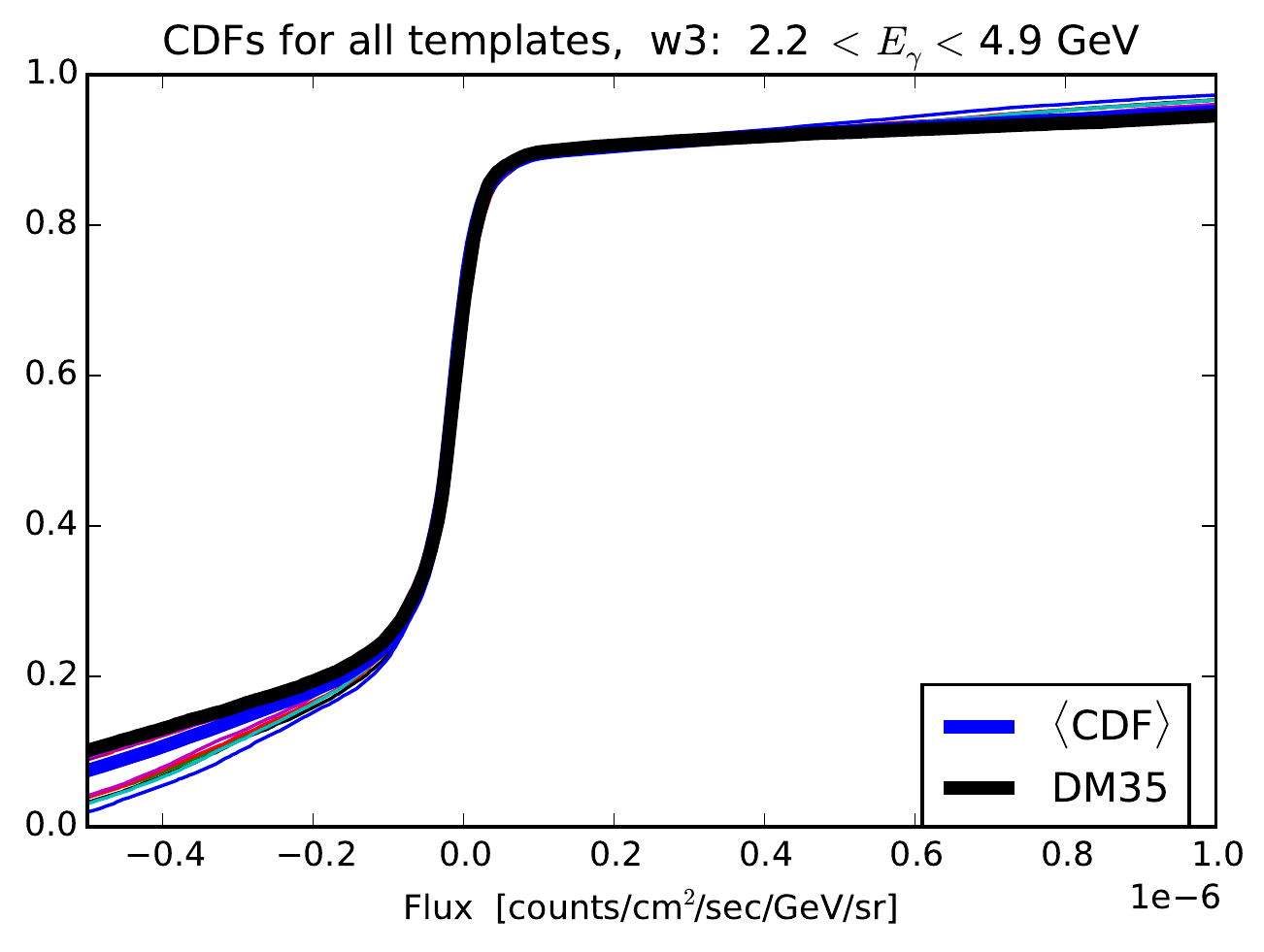}~~~
\includegraphics[width=.3\textwidth]{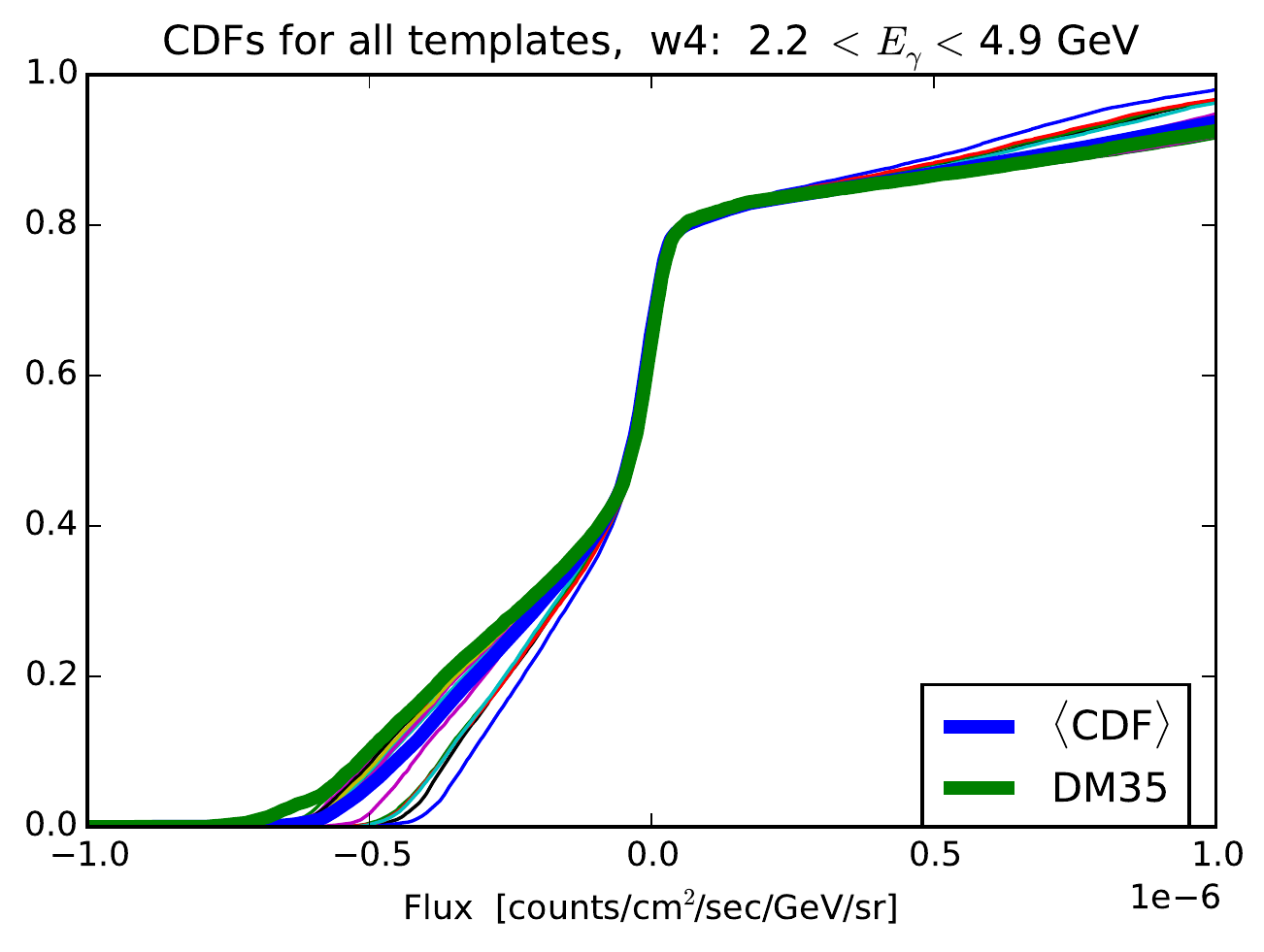}~~~
\includegraphics[width=.3\textwidth]{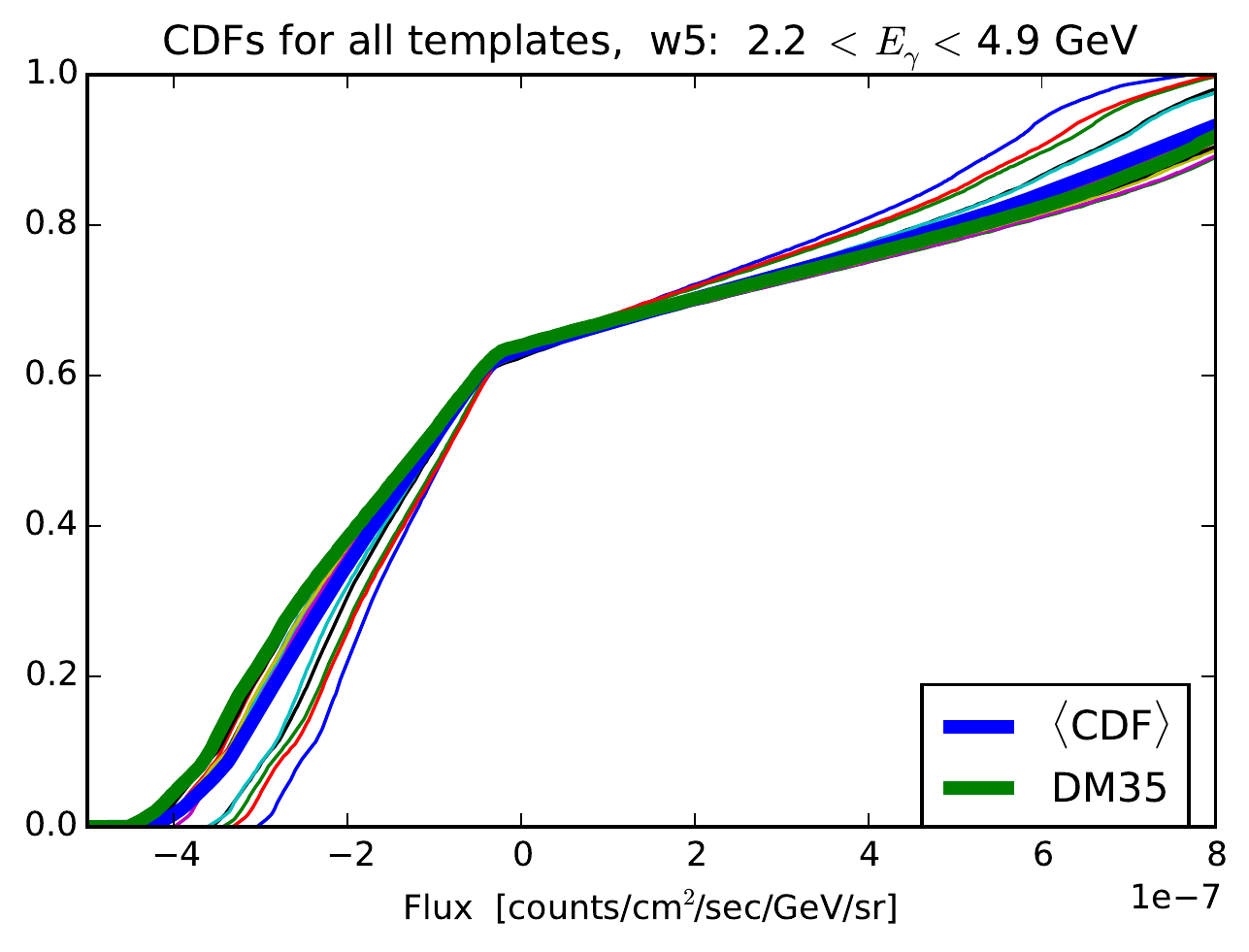}
\caption{Histograms and CDFs of flux per pixel for $\langle\cdf\rangle$ (thick blue) and for a 35 GeV dark matter signal with $M\cdot \langle \sigma v \rangle = 30 \times 10^{-26}\cm^3\!\!/$s (see \Tab{tab:SimulationsTab}), for wavelet levels 3 through 5. We color the dark matter histogram and CDF green if $W_j \neq0$ as defined in \Eq{wjgreater}.}
\label{signal hists}
\end{center}
\end{figure}

In \Figs{signal hists}{cleaned residual example}, we demonstrate the method by which we will isolate new extended emission from dark matter annihilation that could plausibly explain the \textit{Fermi} Galactic center excess. As summarized in \Tab{tab:SimulationsTab}, the mock data set ``DM35'' includes a signal template that approximately mimics a 35 GeV dark matter particle annihilating to $\bar b b$ pairs with $\langle \sigma v \rangle \simeq 10^{-26}\cm^3\!\!/$s; for ease of interpretation, we use a simple power-law photon spectrum, described in \App{app:Simul}, with photon multiplicity per annihilation times the annihilation cross section $M\cdot \langle \sigma v \rangle =  30\times 10^{-26}\cm^3\!\!/$s.  In \Fig{signal hists} we compare a signal histogram and its associated CDF with $\langle\cdf\rangle$, which is the average of the CDFs of the background templates. We show this comparison at three wavelet levels, covering angular scales from roughly $3^\circ$ to $22^\circ$. This demonstrates the method outlined in \Sec{sec:Cleaning Maps}: if the CDF of the signal is sufficiently different from $\langle\cdf\rangle$, the wavelet level is retained for further analysis. We see that $w_4$ and $w_5$ (corresponding to angular scales from approximately $6^\circ$ to $22^\circ$) differ significantly from the background expectation and exceed the significance threshold outlined in \Sec{sec:Cleaning Maps}, while $w_3$ (at smaller angular scales, from $3^\circ$ to $6^\circ$) has a CDF that does not score high enough on the KS test to exceed this threshold. This can be seen in the top row by noticing that the tail of the signal histogram extends considerably farther than the background histogram for $w_4$ and $w_5$, but not $w_3$. Equivalently, in the lower row we see that the maximal distance between the signal CDF and the average CDF is relatively large at the higher levels. This does not mean that the signal in the dark matter dataset is restricted to $w_4$ and above. In fact, the signal is also present on $w_3$ and lower, but the systematic uncertainties are more pronounced there. For those levels, the signal does not rise above the thresholding procedure outlined in \Sec{sec:Cleaning Maps}.

\begin{figure}[tbp]
\begin{center}
\includegraphics[width=\textwidth]{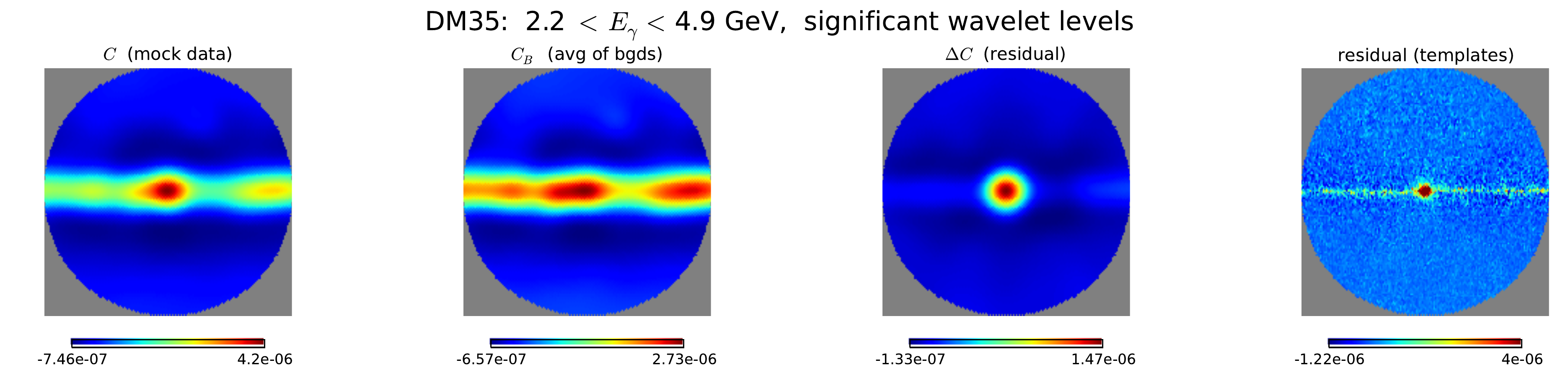}
\caption{$C$ as defined in \Eq{clean} for a 35 GeV dark matter signal with $M\cdot \langle \sigma v \rangle = 30 \times 10^{-26}\cm^3\!\!/$s (left); $C_B$, the sum of the average background over the same wavelet levels, (left middle); and their difference, $\Delta C$, for the mock data set (right middle). We compare this to the residual using a simple subtraction of the average template from the signal (right). The region shown is within an opening angle of $30^\circ$ of the Galactic center.}
\label{cleaned residual example}
\end{center}
\end{figure}

In \Fig{cleaned residual example} we compare the output of our method with the residual from a template-only method. The left panels show $C$ and $C_B$ for the signal analyzed in \Fig{signal hists}, while the third panel shows $\Delta C$. When we compare this to the rightmost panel, it is evident that the wavelet-cleaned map offers a clearer view of the excess. This demonstrates the advantage of our method: without inserting any signal information in advance (neither regarding the morphology nor the angular extent), we extract a clear residual image of an excess emission component extending to relatively high latitudes. The extent of the emission is obscured with a map-based subtraction method.

\begin{figure}[tbp]
\begin{center}
\includegraphics[width=\textwidth]{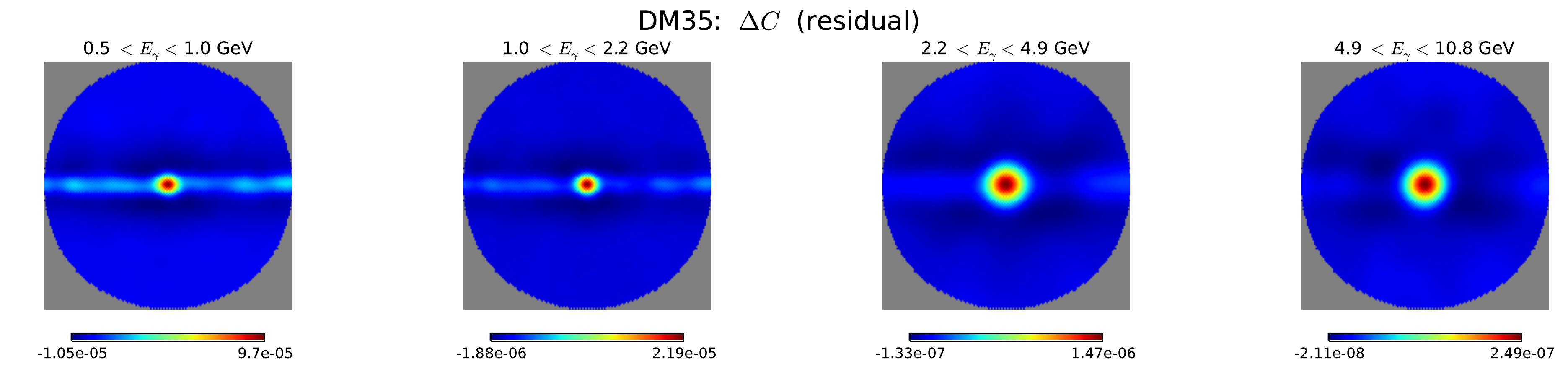}
\includegraphics[width=\textwidth]{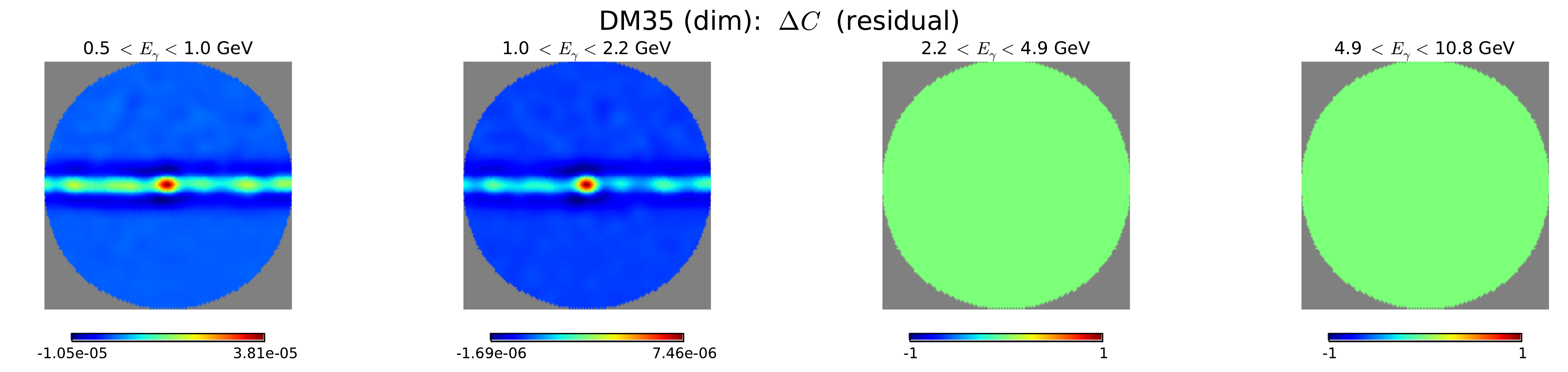}
\caption{$\Delta C$ as defined in \Eq{resid} for a 35 GeV dark matter signal with $M\cdot \langle \sigma v \rangle = 30 \times 10^{-26}\cm^3\!\!/$s (top) and with $M\cdot \langle \sigma v \rangle = 9 \times 10^{-26}\cm^3\!\!/$s (bottom). The region shown is within an opening angle of $30^\circ$ of the Galactic center.}
\label{DM threshold}
\end{center}
\end{figure}

Like any method, thresholding based on wavelet coefficients has limits. We are not able to reconstruct signals generated by arbitrarily dim excesses, as we eventually hit a sensitivity threshold. In \Fig{DM threshold} we show $\Delta C$ for a 35 GeV dark matter signal in various energy bins, varying only the annihilation cross section in the mock data. For the large annihilation cross section of the ``DM35'' data set (top) we see a clear signal. When the annihilation cross section is decreased and the signal is commensurately less bright, as is the case for the ``DM35 (dim)'' simulation (bottom), which has $M\cdot \langle \sigma v \rangle = 9 \times 10^{-26}\cm^3\!\!/$s, we see a residual with significant wavelet levels only in some energy bins. Furthermore, the signal we do find is less obviously spherically symmetric, reflecting more background contamination.

Lower cross sections make the reconstruction of the original mock signal more difficult. Given the non-parametric nature of our thresholding procedure, assigning a value for a threshold is not straightforward. Because the Kolmogorov--Smirnov test values of the backgrounds and the signal both vary with energy, it is possible for a signal to give a nonzero residual in an energy bin but not in its adjacent bins. For values near the detection threshold, the range of background models considered and even different instantiations of the Poisson noise can affect the precise luminosity for which a signal residual will be nonzero. For 35 GeV dark matter, direct investigation reveals that this threshold is around $\bL M\cdot \langle \sigma v \rangle  \bR^{m_\chi=35\gev}_{\rm thresh} \sim 10 \times 10^{-26}\cm^3\!\!/$s, while a similar analysis with 10 GeV dark matter gives $\bL M\cdot \langle \sigma v \rangle \bR^{m_\chi=10\gev}_{\rm thresh} \sim 0.3 \times 10^{-26}\cm^3\!\!/$s. Since our approach incorporates the systematic uncertainties at such a low level, this kind of analysis is of more practical interest for bright signals that exceed the systematic uncertainties.

\begin{figure}[tbp]
\begin{center}
\includegraphics[width=\textwidth]{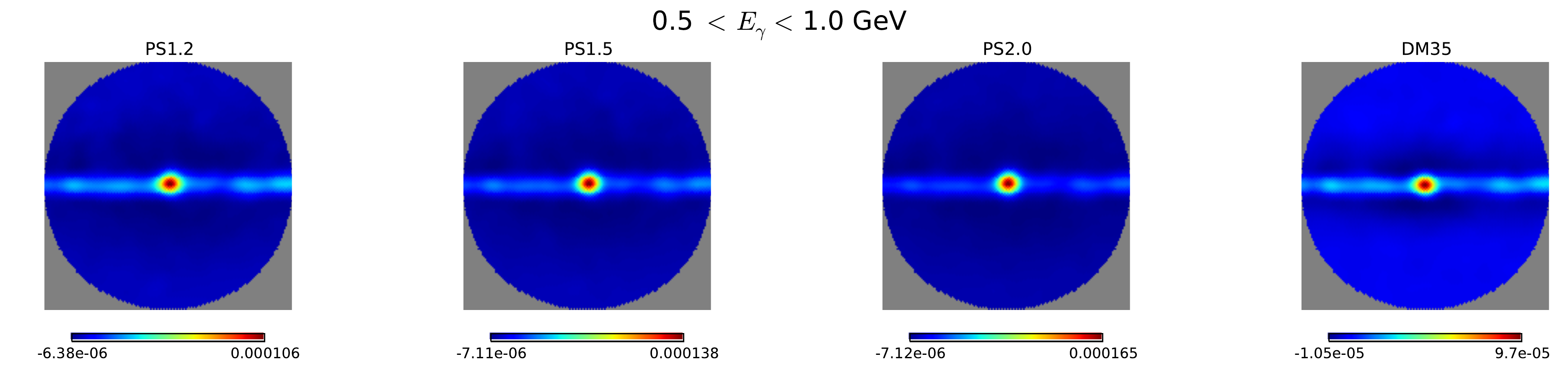}
\includegraphics[width=\textwidth]{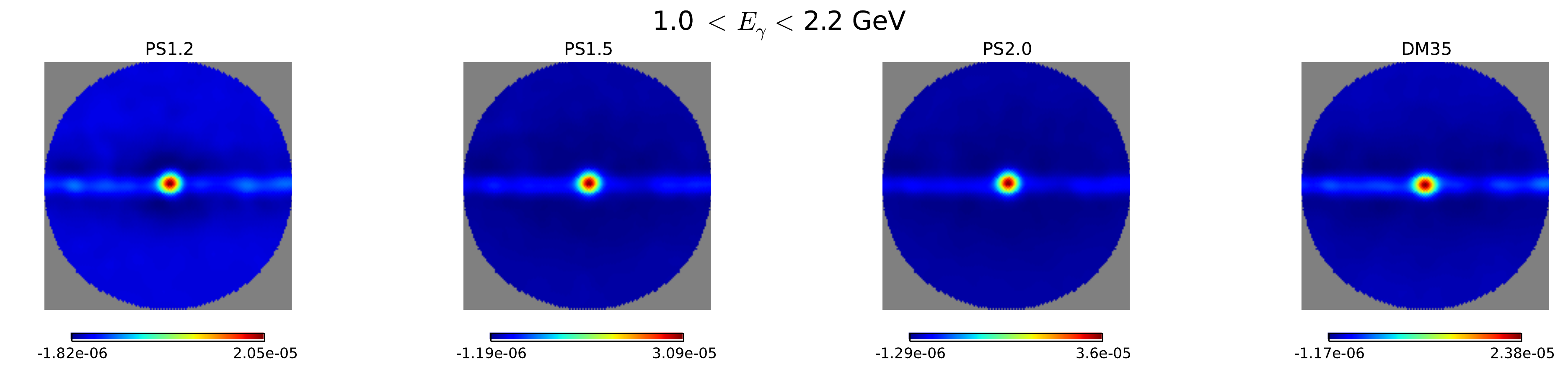}
\includegraphics[width=\textwidth]{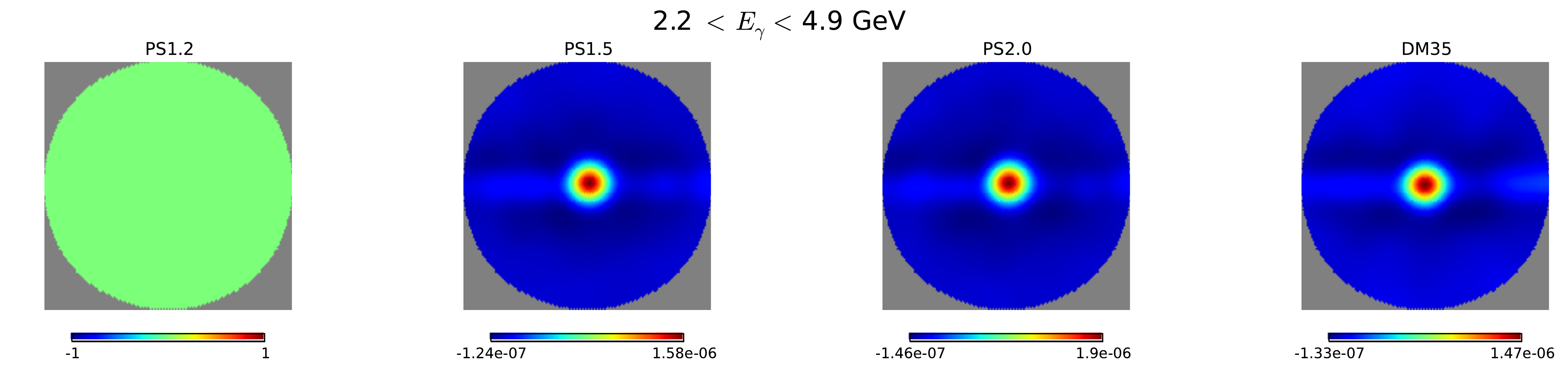}
\includegraphics[width=\textwidth]{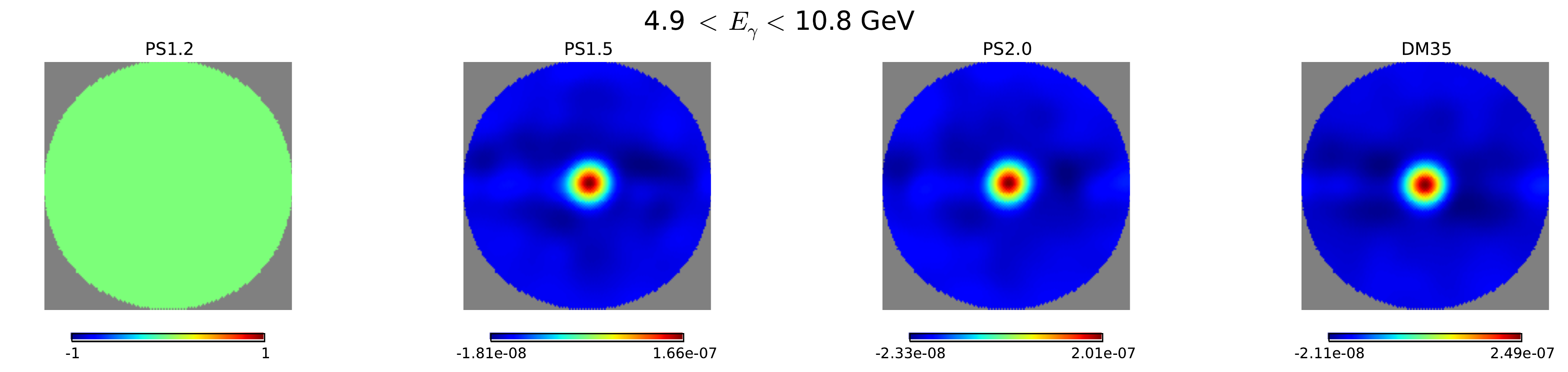}
\caption{$\Delta C$ as defined in \Eq{resid} for a 35 GeV dark matter signal with $M\cdot \langle \sigma v \rangle = 30 \times 10^{-26}\cm^3\!\!/$s (right) compared to point source signals with $\alpha_L = 1.2$ (left), 1.5 (left middle), and 2 (right middle). The region shown is within an opening angle of $30^\circ$ of the Galactic center.}
\label{compare PS}
\end{center}
\end{figure}

One of our primary motivations in this work is to provide a template-free method for extracting excess signal components. Being able to differentiate between candidate signals is one obvious goal that is currently possible only by inserting the templates at the outset and doing different analyses with, possibly, different numbers of free parameters. To this end, we examine the ability of our wavelet-based method to differentiate between dark matter annihilation and point source emission with the same {\it total} photon flux and a relatively low luminosity cutoff $L_c = 1.0 \times 10^{34}$ erg/s, keeping a single free parameter $\alpha_L$ which describes the slope of the point source luminosity function (see \App{app:Simul} for details). This slope parameter is poorly constrained, but most models fix it to be around $\alpha_L \sim 1.5$ \cite{Lee:2014mza,Lee:2015fea,Petrovic:2014xra,Cholis:2014noa}, so for our simulations we adopt the range $1.2 \leq \alpha_L \leq 2$. For a fixed total luminosity, smaller (larger) values of $\alpha_L$ correspond to more (fewer) sources near the cutoff of the luminosity function. As we see in \Fig{compare PS}, for very low $\alpha_L$ we remove point source emission more effectively compared to dark matter emission, as evidenced by the number of vanishing residual images. This is as expected since low $\alpha_L$ demands fewer total sources, a larger portion of which must be near the detection threshold; at high energy (when the total flux is decreasing) these individual bright sources are necessarily more isolated from one another. Our wavelet analysis removes them because their contribution is restricted to low wavelet scales where background variations are large, and no residual is seen in $\Delta C$. In contrast, for large $\alpha_L$, corresponding to a large number of faint sources, the emission morphology is smoother and more extended so that the wavelet thresholding procedure is less effective at removing the structures that lead to the emission. Thus, point sources near or above threshold can be removed by a wavelet analysis, while point sources well below threshold are successful at mimicking smooth extended emission.

\subsection{\textit{Fermi} Bubbles}

Because of their ability to find structures with support on different angular scales, wavelets are naturally suited to investigations of the \textit{Fermi} Bubbles. The \textit{Fermi} Bubbles are extended gamma-ray-bright lobes roughly localized within $|\ell|<30^\circ$ and $|b|<50^\circ$ \cite{Su:2010qj}. The Bubbles have sharp edges that appear to be as bright or brighter than their interior and a hard energy spectrum that allows them to be separated by eye from the other emission components at high latitudes and above few GeV.

In our analysis, we use the same set of candidate background models discussed in \Sec{sec:Diffuse Templates} to parameterize the background uncertainties. The mock data we use is described in Table~\ref{tab:SimulationsTab}. We insert Bubbles with an exactly flat morphology within the boundaries described in \cite{Su:2010qj}, with spectral information from \cite{Fermi-LAT:2014sfa} (including a higher-statistics update with extension of the energy spectrum to higher energies). After extracting wavelet coefficients using the IUWTS described in \Sec{sec:Wavelets}, we perform the cleaning procedure described in \Sec{sec:Cleaning Maps} in a region with an opening angle of $50^\circ$ from the Galactic center, masking the Galactic disk below latitudes $\mL b \mR \geq 3^\circ$. As in \cite{Dobler:2009xz, Su:2010qj,Fermi-LAT:2014sfa}, we choose to mask the Galactic disk because the low-latitude uncertainties are substantially bigger than the high-latitude uncertainties, and the KS test (which includes no spatial information) is unnecessarily hampered in finding high-latitude emission if all disk-associated uncertainties are included.  As discussed above, the IUWTS utilized here preferentially selects isotropic structures from an image. In the interest of tracing the location of the Bubble edges to low latitudes, a wavelet transform that breaks isotropy may be justified in the interest of overcoming the large diffuse uncertainties along the Galactic disk. Although such an investigation is of substantial interest, it is beyond the scope of this work, and we simply retain the IUWTS here for high-latitude searches.

\begin{figure}[t]
\begin{center}
\includegraphics[width=\textwidth]{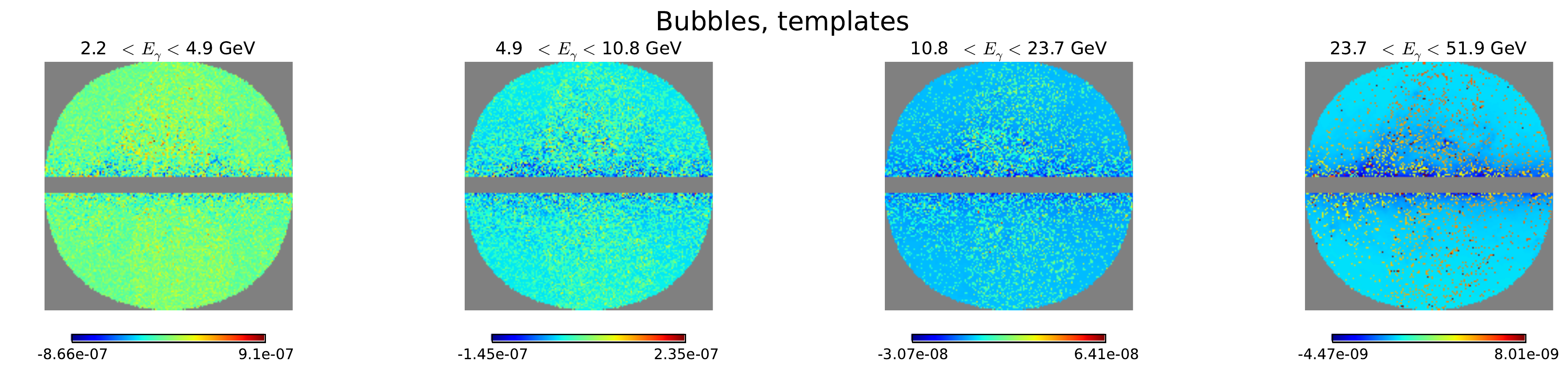}
\includegraphics[width=\textwidth]{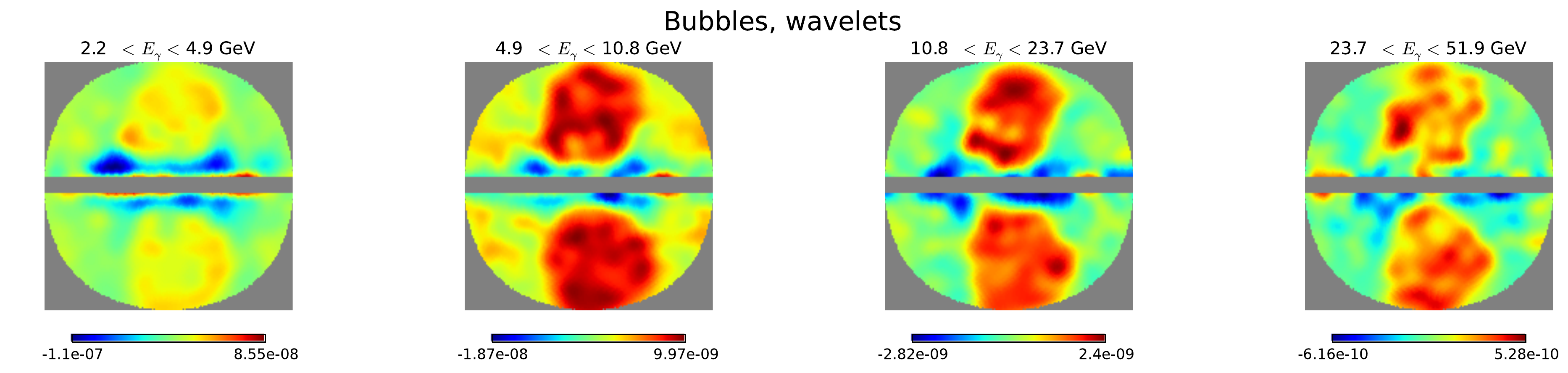}
\caption{A template-based approach for finding the Fermi Bubbles (top) as compared to the cleaned-residual method described here, with $\Delta C$ as defined in \Eq{resid} (bottom). The region shown is within an opening angle of $50^\circ$ of the Galactic center, masking the Galactic disk at $3^\circ$.}
\label{Bubble fig}
\end{center}
\end{figure}

In \Fig{Bubble fig} we display the results of this cleaned analysis with the IUWTS. We see that wavelets find the interior of the Bubbles with remarkable ease. We retain considerable information about their angular extent and energy spectrum using the algorithm described in \Sec{sec:Cleaning Maps}. We compare these results to an equivalent analysis without utilizing wavelets. The difference is stark: the wavelets clearly reveal the extent of both the north and south Bubble. However, the wavelet-based approach reveals specious hotspots within the Bubble geometries. Because the injected mock data had perfectly flat Bubble morphologies, these hotspots probably come from a combination of statistical noise and variations of the background templates at high latitudes.

\section{Discussion and Conclusions}
\label{sec:Conclusion}

Wavelets are a powerful tool for data analysis. They provide information on conjugate features of a data set, allowing simultaneous access to the size and location of the constituent structures of a composite signal. This provides a natural mechanism for separating astrophysical uncertainties at different scales, since the backgrounds can be well constrained on one angular scale but poorly understood on another. Comparing the wavelet transforms of different background models then provides a natural measure for the significance of any deviation in data, and allows a convenient thresholding device for cleaning an image to reveal interesting structures.

The main output of our analysis is a cleaned residual image $\Delta C$ that allows a characterization of a given data set without reference to a specific background model. Without assuming the existence of a new signal component, we can still determine the characteristics of a data set that contains a new signal that is incompatible with the systematic background uncertainties.  

Specifically, we implemented the isotropic undecimated wavelet transform on the sphere \cite{Starck:2005fb} and showed how it could be used to extract information about extended emission towards the Galactic center. Using mock data, in order to have full control and to be able to better assess the capabilities of this approach, we investigated the utility of wavelets to understand gamma-ray excesses due to dark matter annihilation, the \textit{Fermi} bubbles, and a collection of point sources.  
The wavelet approach is able to find a dark matter contribution, of the type previously proposed to explain the Galactic center excess, down to annihilation cross sections below the thermal rate.  The \textit{Fermi} bubbles are easily found across a broad energy range.
In comparing dark matter annihilation to various point source populations, we showed how the wavelet transform can potentially discriminate between these distributions, if the luminosity function for the point sources is not too steep, without assuming anything about the spatial morphology of the signal.

The biggest next step is to adapt this methodology to actual \textit{Fermi} data \cite{futurepaper}. We will be able to perform model-independent tests of the morphology of the Galactic center excess. The approach outlined here is powerful enough to work with relatively small energy bins that may allow a simultaneous spectral analysis of this excess. It may also be interesting to adapt this analysis to more general mother wavelets that identify different signal features. For example, wavelet transforms with preferred axes like the ridgelet or curvelet transforms \cite{Starck:2005fb} may be of value for investigating the edges of the \textit{Fermi} Bubbles. Using advanced wavelet methods like the multiscale resolution analysis could provide interesting insight into the large-scale structure of the Bubbles. Other extended sources in the gamma-ray sky that likely arise from astrophysical processes along our line of sight across the disk \cite{blobs} will also be exciting to probe with a wavelet-based approach. Characterizing their origin and composition while remaining agnostic about their associated cosmic ray sources will be an exciting application of the wavelet-based signal identification we have described here.

Wavelets are promising tools beyond the study of diffuse gamma-ray emission. For example, further understanding the earliest structures in the universe may be possible with wavelet-based image processing of microwave data. Clusters which are identified using multiple frequencies via the spectral shift from the Sunyaev-Zel'dovich effect can have morphological follow-up studies from ground-based cameras with better angular resolution. Finding a continued pattern of Sunyaev-Zel'dovich clusters in the large-scale microwave data on patches of smaller size will potentially increase the number of targets and improve the chances of learning about these structures. Multi-messenger astronomy can also be enhanced with wavelet-based techniques. Given pointing information in one channel, wavelets can potentially clean images in other bands to look for interesting underlying structures.

Wavelet decompositions offer a powerful approach to data-driven understanding of signals, even in the face of significant systematic uncertainties. We have demonstrated this with simulations of the gamma-ray sky, and we have shown how to recover new extended emission when uncertainties dominate at smaller angular scales. Applying these techniques to real data promises to shed light on some of the most interesting structures in the sky.

~\\
\noindent
{\it Acknowledgements}: We would like to thank T.~Brandt, E.~Charles, V.~Gluscevic, D.~Hooper, M.~Lisanti, P.~Luthy, M.~Luty, J.~Mardon, S.~Murgia, B.~Safdi, N.~Shaviv, T.~Slatyer, J.~Thaler, C.~Weniger, W.~Xue, and G.~Zweig for useful discussions. SDM also thanks the CTP at MIT and the astronomy group at IAS for feedback and hospitality while this work was underway. The work of SDM was performed in part at the Aspen Center for Physics, which is supported by National Science Foundation grant PHY-1066293, and at the Galileo Galilei Insitute. SDM is supported by NSF PHY1316617. IC is supported by NASA NNX15AB18G. PJF and IC would like to thank the Korea Institute for Advanced Study for their hospitality provided during the completion of this work.  This work has made use of the {\tt SciPy} \cite{scipy}, {\tt iPython} \cite{ipython}, {\tt HEALPix} \cite{Gorski:2004by}, and {\tt healpy} packages.

\appendix
\section{Statistical Error Bars for Wavelet Maps}
\label{stat unc}

The smoothed maps $c_j$ are defined in terms of the spherical harmonic coefficients of the maps, the $a_\lm$, and the smoothing function, $\hat \phi_j(\ell,0)$, which has implicit cutoff $\ell_c=\ell_{\rm max}/2^j$. The variance of the $c_j$ is different at each point on the sky; $\Omega_{\rm p}$ is used to emphasize that this solid angle refers to a particular direction. We define the smoothed maps as:
\alg{ \label{cj1}
c_{j+1}(\Omega_{\rm p};E) &= \sum_{\ell, m} \sqrt{\frac{4\pi}{2\ell+1}} \hat \phi_j(\ell,0)a_\lm(E) Y_\lm(\Omega_{\rm p}) 
}
This definition and the following discussion are trivially extended to any linear function on the $\hat \phi_j$, like the wavelet maps $w_j \sim \phi_{j-1}-\phi_j$. We expand \Eq{cj1} to get:
\alg{
\langle\mL c_{j+1}(\Omega_{\rm p};E) \mR^2 \rangle &= 4\pi  \sum_{\ell,\dots} \frac{\hat \phi_j(\ell,0) \hat \phi_j(\ell',0)}{\sqrt{(2\ell+1)(2\ell'+1)}} Y_\lm(\Omega_{\rm p}) Y_{\ell'm'}(\Omega_{\rm p})
\times \\ &\times \int_{\tilde \Omega,\dot \Omega} Y_\lm(\tilde \Omega) Y_{\ell'm'}(\dot \Omega) \left< \bL \bar f(\tilde \Omega;E) + \Delta f(\tilde \Omega;E) \bR \bL \bar f(\dot \Omega;E) + \Delta f(\dot \Omega;E) \bR \right>
,}
The integrals over $\tilde \Omega,\dot \Omega$ are integrals over the entire solid angle. We are interested in the variance, and after some standard simplifications (e.g., assuming independent errors for each pixel) we are left with:
\alg{\label{cj2}
\Delta c_{j+1}^2(\Omega_{\rm p};E) &= \int_{\tilde \Omega} \left<  \Delta f(\tilde \Omega;E)^2 \right> \bL \sum_\ell \sqrt{ \frac{2\ell+1}{4\pi} }  \hat \phi_j(\ell,0)  P_\ell(\Omega_{\rm p} \cdot \tilde \Omega) \bR^2  ,
}
where we have used the addition theorem, $\sum_m Y_\lm (\Omega_{\rm p}) Y_\lm(\tilde \Omega) = (2\ell+1)P_\ell(\Omega_{\rm p} \cdot \tilde \Omega)/4\pi$. Since we are interested in the error bars for pixels on a binned sky, we take the discrete limit of \Eq{cj2} with $\int_{\tilde \Omega} d \tilde \Omega \to \frac{4\pi}{N_{\rm pix}} \sum_{\rm pix}$. We find:
\alg{
\Delta c_{j+1}^2(\Omega_{\rm p};E) & =  \frac1{N_{\rm pix}} \sum_{k \in \Omega_{\rm orig}} \frac{\Phi^M_\gamma(k;E)}{\ep(k;E)}   \bL \sum_\ell  \sqrt{2\ell+1} \hat \phi_j(\ell,0)  P_\ell(\Omega_{\rm p} \cdot \Omega_k) \bR^2 .
}
The summation goes over all pixels in the slice of sky $\Omega_{\rm orig}$ on which we did the initial spherical harmonic decomposition (which in practice will always be $4\pi$); $N_{\rm pix}$ is the number of pixels in that slice; and we specialize to the case that the original map is an energy-dependent set of flux maps, such that for each pixel $k$ we define $\Phi^M_\gamma(k;E)$ as the flux and $\ep(k;E)$ as the exposure times energy. Note that both  $\Phi^M(k;E)$ and $\ep(k;E)$ are energy dependent, but whereas $\Phi^M(k;E)$ is map dependent, $\ep(k;E)$ is map independent. For ease of computation, we rewrite as:
\alg{
\Delta c_{j+1}^2(\Omega_{\rm p};E) &= \frac1{N_{\rm pix}} \sum_{k \in \Omega_{\rm orig}} \frac{\Phi^M_\gamma(k;E)}{\ep(k;E)} \mathbb M_{j,{\rm p}}(k) ,
}
where
\beq
\mathbb M_{j,{\rm p}}(k) = \bL \sum_\ell  \sqrt{2\ell+1} \hat \phi_j(\ell,0) P_\ell(\Omega_{\rm p} \cdot \Omega_k) \bR^2.
\eeq
Because $\mathbb M_{j{\rm,p}}$ is the square of a Legendre series, it can be computed quickly once the $\Omega_{\rm p} \cdot \Omega_k$ are obtained.

\section{Simulated Data}
\label{app:Simul}

Here we describe the simulated data.

For the case of a dark matter annihilation signal, we assume an NFW profile:
\beq
\rho(r) = \rho_{0} \cdot \frac{R_{c}}{r}\cdot\frac{1}{(1 + \frac{r}{R_{c}})^{2}}
\eeq
with $\rho_{0}$=0.345 GeV/cm$^3$ and $R_{c}$ = 20 kpc, giving a local value at $r$= 8.5 kpc of 0.4 GeV/cm$^3$. The dark matter flux is given by:
\beq
\Phi_{\rm \gamma, DM}(E) = \frac{1}{2} \frac{1}{m_{\chi}^{2}}\frac{\langle \sigma v \rangle}{4 \pi} \int d\ell \rho(r)^{2} \frac{dN}{dE}\cdot M,
\eeq
where $m_{\chi}$ is the dark matter mass, $\langle \sigma v \rangle$ is the cross-section, $M$ is the photon multiplicity per annihilation event, $dN/dE$ is the differential spectrum, and the factor of $1/2$ comes from assuming a self-conjugate dark matter particle. Rather than restricting to a particular particle physics model, we assume a simple functional form for $dN/dE$:
\beq
\frac{dN}{dE}(E,\alpha_E,E_c) = N(\alpha_E) \pL \frac E{100 \mev} \pR^{-\alpha_E} e^{-E/E_c },
\eeq
with $E_c \sim m_\chi/5$ and $\alpha_E \sim 1-2$ depending on the dark matter mass. These choices are motivated by fits to photon spectra from dark matter annihilations to heavy quarks and gauge bosons. The norm $N(\alpha_E)$ is chosen such that $\int_{\rm 1 MeV}^{m_{\chi}} dE \frac{dN}{dE}$ = 1.

We generate point sources distributed with a $\rho_{r} \propto r^{-2}$ profile in the inner galaxy using a range of luminosity functions. Our luminosity functions are parameterized as:
\beq
N(L,\alpha_L, L_c) = c(\alpha_L) \times L^{-\alpha_L} e^{-L/L_c},
\eeq
with a cutoff $L_c$ and a normalization $c(\alpha_L)$ that is fixed by holding constant the integrated luminosity $\int d L \, L\,N(L,\alpha_L) \simeq 10^{37}$ erg/s. The gamma-ray spectra of those point sources resemble those from MSP observations \cite{Cholis:2014noa}. The spectral parameter $\alpha_L$ is allowed to vary between the wide range of $1.2\leq \alpha \leq 2$.  We fix $L_c$ to $10^{34}$ erg/s.

\bibliographystyle{JHEP}
\bibliography{wavelets}

\providecommand{\href}[2]{#2}\begingroup\raggedright\begin{thebibliography}{100}

\bibitem{Buckley:2013bha}
J.~Buckley {\em et.~al.}, {\it {Working Group Report: WIMP Dark Matter Indirect
  Detection}},  in {\em {Community Summer Study 2013: Snowmass on the
  Mississippi (Cs$S^2$013) Minneapolis, Mn, Usa, July 29-August 6, 2013}},
  2013.
\newblock \href{http://xxx.lanl.gov/abs/1310.7040}{{\tt arXiv:1310.7040}}.

\bibitem{Cushman:2013zza}
P.~Cushman {\em et.~al.}, {\it {Working Group Report: WIMP Dark Matter Direct
  Detection}},  in {\em {Community Summer Study 2013: Snowmass on the
  Mississippi (Cs$S^2$013) Minneapolis, Mn, Usa, July 29-August 6, 2013}},
  2013.
\newblock \href{http://xxx.lanl.gov/abs/1310.8327}{{\tt arXiv:1310.8327}}.

\bibitem{Abbott:2005bi}
{\bf Dark Energy Survey} Collaboration, T.~Abbott {\em et.~al.}, {\it {The dark
  energy survey}},  \href{http://xxx.lanl.gov/abs/astro-ph/0510346}{{\tt
  astro-ph/0510346}}.

\bibitem{Levi:2013gra}
{\bf DESI} Collaboration, M.~Levi {\em et.~al.}, {\it {The Desi Experiment, a
  Whitepaper for Snowmass 2013}},
  \href{http://xxx.lanl.gov/abs/1308.0847}{{\tt arXiv:1308.0847}}.

\bibitem{Pancino:2012aa}
{\bf Gaia-ESO Survey} Collaboration, E.~Pancino, {\it {The Gaia-Eso Survey
  Astrophysical Calibration}},  \href{http://xxx.lanl.gov/abs/1206.6291}{{\tt
  arXiv:1206.6291}}.

\bibitem{futurepaper}
I.~Cholis, P.~J. Fox, and S.~D. McDermott. {\it Work in Progress}.

\bibitem{1992tlw..conf.....D}
I.~Daubechies, {\em CBMS-NSF regional conference series in applied
  mathematics}.
\newblock Society for Industrial and Applied Mathematics, 1992.

\bibitem{AstroImageBook}
J.-L. Starck and F.~Murtaugh., {\em Astronomical Image and Data Analysis}.
\newblock Astronomy and Astrophysics Library, Springer, New York, 2nd~ed.,
  2006.

\bibitem{Starck:2005fb}
J.-L. Starck, Y.~Moudden, P.~Abrial, and M.~Nguyen, {\it {Wavelets, Ridgelets
  and Curvelets on the Sphere}},  {\em Astron. Astrophys.} {\bf 446} (2006)
  1191, [\href{http://xxx.lanl.gov/abs/astro-ph/0509883}{{\tt
  astro-ph/0509883}}].

\bibitem{Schmitt:2010vf}
J.~Schmitt, J.~L. Starck, J.~M. Casandjian, J.~Fadili, and I.~Grenier, {\it
  {Poisson Denoising on the Sphere: Application to the Fermi Gamma Ray Space
  Telescope}},  {\em Astron. Astrophys.} {\bf 517} (2010) A26,
  [\href{http://xxx.lanl.gov/abs/1003.5613}{{\tt arXiv:1003.5613}}].

\bibitem{Rentala:2014bxa}
V.~Rentala, W.~Shepherd, and T.~M.~P. Tait, {\it {Tagging Boosted Ws with
  Wavelets}},  {\em JHEP} {\bf 08} (2014) 042,
  [\href{http://xxx.lanl.gov/abs/1404.1929}{{\tt arXiv:1404.1929}}].

\bibitem{Monk:2014uza}
J.~W. Monk, {\it {Wavelet Analysis: Event De-Noising, Shower Evolution and Jet
  Substructure without Jets}},  \href{http://xxx.lanl.gov/abs/1405.5008}{{\tt
  arXiv:1405.5008}}.

\bibitem{keinertbook}
F.~Keinert, {\em Wavelets and Multiwavelets}.
\newblock Studies in Advanced Mathematics, CRC Press, 2003.

\bibitem{Chan:2015jiv}
J.~Y.~H. Chan, B.~Leistedt, T.~D. Kitching, and J.~D. McEwen, {\it
  {Second-Generation Curvelets on the Sphere}},
  \href{http://xxx.lanl.gov/abs/1511.0557}{{\tt arXiv:1511.0557}}.

\bibitem{1999APh....11..277G}
N.~Gehrels and P.~Michelson, {\it Glast facility science team},  {\em
  Astroparticle Physics} {\bf 11} (1999), no.~277.

\bibitem{Dobler:2009xz}
G.~Dobler, D.~P. Finkbeiner, I.~Cholis, T.~R. Slatyer, and N.~Weiner, {\it {The
  Fermi Haze: a Gamma-Ray Counterpart to the Microwave Haze}},  {\em Astrophys.
  J.} {\bf 717} (2010) 825--842, [\href{http://xxx.lanl.gov/abs/0910.4583}{{\tt
  arXiv:0910.4583}}].

\bibitem{Su:2010qj}
M.~Su, T.~R. Slatyer, and D.~P. Finkbeiner, {\it {Giant Gamma-Ray Bubbles from
  Fermi-Lat: Agn Activity Or Bipolar Galactic Wind?}},  {\em Astrophys. J.}
  {\bf 724} (2010) 1044--1082, [\href{http://xxx.lanl.gov/abs/1005.5480}{{\tt
  arXiv:1005.5480}}].

\bibitem{Fermi-LAT:2014sfa}
{\bf Fermi-LAT} Collaboration, M.~Ackermann {\em et.~al.}, {\it {The Spectrum
  and Morphology of the $Fermi$ Bubbles}},  {\em Astrophys. J.} {\bf 793}
  (2014), no.~1 64, [\href{http://xxx.lanl.gov/abs/1407.7905}{{\tt
  arXiv:1407.7905}}].

\bibitem{Goodenough:2009gk}
L.~Goodenough and D.~Hooper, {\it {Possible Evidence for Dark Matter
  Annihilation in the Inner Milky Way from the Fermi Gamma Ray Space
  Telescope}},  \href{http://xxx.lanl.gov/abs/0910.2998}{{\tt
  arXiv:0910.2998}}.

\bibitem{Hooper:2010mq}
D.~Hooper and L.~Goodenough, {\it {Dark Matter Annihilation in the Galactic
  Center as Seen by the Fermi Gamma Ray Space Telescope}},
  \href{http://xxx.lanl.gov/abs/1010.2752}{{\tt arXiv:1010.2752}}.

\bibitem{Hooper:2011ti}
D.~Hooper and T.~Linden, {\it {On the Origin of the Gamma Rays from the
  Galactic Center}},  {\em Phys.Rev.} {\bf D84} (2011) 123005,
  [\href{http://xxx.lanl.gov/abs/1110.0006}{{\tt arXiv:1110.0006}}].

\bibitem{Abazajian:2012pn}
K.~N. Abazajian and M.~Kaplinghat, {\it {Detection of a Gamma-Ray Source in the
  Galactic Center Consistent with Extended Emission from Dark Matter
  Annihilation and Concentrated Astrophysical Emission}},  {\em Phys.Rev.} {\bf
  D86} (2012) 083511, [\href{http://xxx.lanl.gov/abs/1207.6047}{{\tt
  arXiv:1207.6047}}].

\bibitem{Hooper:2013rwa}
D.~Hooper and T.~R. Slatyer, {\it {Two Emission Mechanisms in the Fermi
  Bubbles: a Possible Signal of Annihilating Dark Matter}},  {\em Phys.Dark
  Univ.} {\bf 2} (2013) 118--138,
  [\href{http://xxx.lanl.gov/abs/1302.6589}{{\tt arXiv:1302.6589}}].

\bibitem{Gordon:2013vta}
C.~Gordon and O.~Macias, {\it {Dark Matter and Pulsar Model Constraints from
  Galactic Center Fermi-Lat Gamma Ray Observations}},  {\em Phys.Rev.} {\bf
  D88} (2013), no.~8 083521, [\href{http://xxx.lanl.gov/abs/1306.5725}{{\tt
  arXiv:1306.5725}}].

\bibitem{Abazajian:2014fta}
K.~N. Abazajian, N.~Canac, S.~Horiuchi, and M.~Kaplinghat, {\it {Astrophysical
  and Dark Matter Interpretations of Extended Gamma-Ray Emission from the
  Galactic Center}},  {\em Phys.Rev.} {\bf D90} (2014), no.~2 023526,
  [\href{http://xxx.lanl.gov/abs/1402.4090}{{\tt arXiv:1402.4090}}].

\bibitem{Daylan:2014rsa}
T.~Daylan, D.~P. Finkbeiner, D.~Hooper, T.~Linden, S.~K.~N. Portillo, {\em
  et.~al.}, {\it {The Characterization of the Gamma-Ray Signal from the Central
  Milky Way: a Compelling Case for Annihilating Dark Matter}},
  \href{http://xxx.lanl.gov/abs/1402.6703}{{\tt arXiv:1402.6703}}.

\bibitem{Calore:2014xka}
F.~Calore, I.~Cholis, and C.~Weniger, {\it {Background Model Systematics for
  the Fermi GeV Excess}},  {\em JCAP} {\bf 1503} (2015) 038,
  [\href{http://xxx.lanl.gov/abs/1409.0042}{{\tt arXiv:1409.0042}}].

\bibitem{TheFermi-LAT:2015kwa}
{\bf Fermi-LAT} Collaboration, M.~Ajello {\em et.~al.}, {\it {Fermi-LAT
  Observations of High-Energy Gamma-Ray Emission Toward the Galactic Center}},
  \href{http://xxx.lanl.gov/abs/1511.0293}{{\tt arXiv:1511.0293}}.

\bibitem{Strong:2007nh}
A.~W. Strong, I.~V. Moskalenko, and V.~S. Ptuskin, {\it {Cosmic-Ray Propagation
  and Interactions in the Galaxy}},  {\em Ann. Rev. Nucl. Part. Sci.} {\bf 57}
  (2007) 285--327, [\href{http://xxx.lanl.gov/abs/astro-ph/0701517}{{\tt
  astro-ph/0701517}}].

\bibitem{Ackermann:2012pya}
{\bf Fermi-LAT} Collaboration, M.~Ackermann {\em et.~al.}, {\it {Fermi-LAT
  Observations of the Diffuse Gamma-Ray Emission: Implications for Cosmic Rays
  and the Interstellar Medium}},  {\em Astrophys. J.} {\bf 750} (2012) 3,
  [\href{http://xxx.lanl.gov/abs/1202.4039}{{\tt arXiv:1202.4039}}].

\bibitem{KS-orig1}
A.~N. Kolmogorov, {\it Sulla determinazione empirica di una legge di
  distributione},  {\em Giornale dell'Istituto Italiano degli Attuari} {\bf 4}
  (1933), no.~83.

\bibitem{KS-orig2}
N.~Smirnov, {\it Table for estimating the goodness of fit of empirical
  distributions},  {\em Ann. Math. Statist.} {\bf 19} (1948) 279--281.

\bibitem{Acero:2015hja}
{\bf Fermi-LAT} Collaboration, F.~Acero {\em et.~al.}, {\it {Fermi Large Area
  Telescope Third Source Catalog}},
  \href{http://xxx.lanl.gov/abs/1501.0200}{{\tt arXiv:1501.0200}}.

\bibitem{Alves:2014yha}
A.~Alves, S.~Profumo, F.~S. Queiroz, and W.~Shepherd, {\it {Effective Field
  Theory Approach to the Galactic Center Gamma-Ray Excess}},  {\em Phys. Rev.}
  {\bf D90} (2014), no.~11 115003,
  [\href{http://xxx.lanl.gov/abs/1403.5027}{{\tt arXiv:1403.5027}}].

\bibitem{Logan:2010nw}
H.~E. Logan, {\it {Dark Matter Annihilation Through a Lepton-Specific Higgs
  Boson}},  {\em Phys. Rev.} {\bf D83} (2011) 035022,
  [\href{http://xxx.lanl.gov/abs/1010.4214}{{\tt arXiv:1010.4214}}].

\bibitem{Boehm:2014hva}
C.~Boehm, M.~J. Dolan, C.~McCabe, M.~Spannowsky, and C.~J. Wallace, {\it
  {Extended gamma-ray emission from Coy Dark Matter}},  {\em JCAP} {\bf 1405}
  (2014) 009, [\href{http://xxx.lanl.gov/abs/1401.6458}{{\tt
  arXiv:1401.6458}}].

\bibitem{Modak:2013jya}
K.~P. Modak, D.~Majumdar, and S.~Rakshit, {\it {A Possible Explanation of Low
  Energy $\gamma$-ray Excess from Galactic Centre and Fermi Bubble by a Dark
  Matter Model with Two Real Scalars}},  {\em JCAP} {\bf 1503} (2015) 011,
  [\href{http://xxx.lanl.gov/abs/1312.7488}{{\tt arXiv:1312.7488}}].

\bibitem{Huang:2013apa}
W.-C. Huang, A.~Urbano, and W.~Xue, {\it {Fermi Bubbles Under Dark Matter
  Scrutiny Part Ii: Particle Physics Analysis}},  {\em JCAP} {\bf 1404} (2014)
  020, [\href{http://xxx.lanl.gov/abs/1310.7609}{{\tt arXiv:1310.7609}}].

\bibitem{Okada:2013bna}
N.~Okada and O.~Seto, {\it {Gamma Ray Emission in Fermi Bubbles and Higgs
  Portal Dark Matter}},  {\em Phys. Rev.} {\bf D89} (2014), no.~4 043525,
  [\href{http://xxx.lanl.gov/abs/1310.5991}{{\tt arXiv:1310.5991}}].

\bibitem{Hagiwara:2013qya}
K.~Hagiwara, S.~Mukhopadhyay, and J.~Nakamura, {\it {10 GeV Neutralino Dark
  Matter and Light Stau in the MSSM}},  {\em Phys. Rev.} {\bf D89} (2014),
  no.~1 015023, [\href{http://xxx.lanl.gov/abs/1308.6738}{{\tt
  arXiv:1308.6738}}].

\bibitem{Buckley:2013sca}
M.~R. Buckley, D.~Hooper, and J.~Kumar, {\it {Phenomenology of Dirac Neutralino
  Dark Matter}},  {\em Phys. Rev.} {\bf D88} (2013) 063532,
  [\href{http://xxx.lanl.gov/abs/1307.3561}{{\tt arXiv:1307.3561}}].

\bibitem{Anchordoqui:2013pta}
L.~A. Anchordoqui and B.~J. Vlcek, {\it {W-Wimp Annihilation as a Source of the
  Fermi Bubbles}},  {\em Phys. Rev.} {\bf D88} (2013) 043513,
  [\href{http://xxx.lanl.gov/abs/1305.4625}{{\tt arXiv:1305.4625}}].

\bibitem{Buckley:2011mm}
M.~R. Buckley, D.~Hooper, and J.~L. Rosner, {\it {A Leptophobic Z' and Dark
  Matter from Grand Unification}},  {\em Phys. Lett.} {\bf B703} (2011)
  343--347, [\href{http://xxx.lanl.gov/abs/1106.3583}{{\tt arXiv:1106.3583}}].

\bibitem{Boucenna:2011hy}
M.~S. Boucenna and S.~Profumo, {\it {Direct and Indirect Singlet Scalar Dark
  Matter Detection in the Lepton-Specific Two-Higgs-Doublet Model}},  {\em
  Phys. Rev.} {\bf D84} (2011) 055011,
  [\href{http://xxx.lanl.gov/abs/1106.3368}{{\tt arXiv:1106.3368}}].

\bibitem{Marshall:2011mm}
G.~Marshall and R.~Primulando, {\it {The Galactic Center Region Gamma Ray
  Excess from a Supersymmetric Leptophilic Higgs Model}},  {\em JHEP} {\bf 05}
  (2011) 026, [\href{http://xxx.lanl.gov/abs/1102.0492}{{\tt
  arXiv:1102.0492}}].

\bibitem{Zhu:2011dz}
G.~Zhu, {\it {Wimpless Dark Matter and the Excess Gamma Rays from the Galactic
  Center}},  {\em Phys. Rev.} {\bf D83} (2011) 076011,
  [\href{http://xxx.lanl.gov/abs/1101.4387}{{\tt arXiv:1101.4387}}].

\bibitem{Buckley:2010ve}
M.~R. Buckley, D.~Hooper, and T.~M.~P. Tait, {\it {Particle Physics
  Implications for Cogent, Dama, and Fermi}},
  \href{http://xxx.lanl.gov/abs/1011.1499}{{\tt arXiv:1011.1499}}.

\bibitem{Berlin:2014tja}
A.~Berlin, D.~Hooper, and S.~D. McDermott, {\it {Simplified Dark Matter Models
  for the Galactic Center Gamma-Ray Excess}},  {\em Phys. Rev.} {\bf D89}
  (2014), no.~11 115022, [\href{http://xxx.lanl.gov/abs/1404.0022}{{\tt
  arXiv:1404.0022}}].

\bibitem{Izaguirre:2014vva}
E.~Izaguirre, G.~Krnjaic, and B.~Shuve, {\it {The Galactic Center Excess from
  the Bottom Up}},  {\em Phys. Rev.} {\bf D90} (2014), no.~5 055002,
  [\href{http://xxx.lanl.gov/abs/1404.2018}{{\tt arXiv:1404.2018}}].

\bibitem{Agrawal:2014una}
P.~Agrawal, B.~Batell, D.~Hooper, and T.~Lin, {\it {Flavored Dark Matter and
  the Galactic Center Gamma-Ray Excess}},  {\em Phys. Rev.} {\bf D90} (2014),
  no.~6 063512, [\href{http://xxx.lanl.gov/abs/1404.1373}{{\tt
  arXiv:1404.1373}}].

\bibitem{Cerdeno:2014cda}
D.~G. Cerde{\~n}o, M.~Peir{\'o}, and S.~Robles, {\it {Low-mass right-handed
  sneutrino dark matter: SuperCDMS and LUX constraints and the Galactic Centre
  gamma-ray excess}},  {\em JCAP} {\bf 1408} (2014) 005,
  [\href{http://xxx.lanl.gov/abs/1404.2572}{{\tt arXiv:1404.2572}}].

\bibitem{Ipek:2014gua}
S.~Ipek, D.~McKeen, and A.~E. Nelson, {\it {A Renormalizable Model for the
  Galactic Center Gamma Ray Excess from Dark Matter Annihilation}},  {\em Phys.
  Rev.} {\bf D90} (2014), no.~5 055021,
  [\href{http://xxx.lanl.gov/abs/1404.3716}{{\tt arXiv:1404.3716}}].

\bibitem{Ghosh:2014pwa}
D.~K. Ghosh, S.~Mondal, and I.~Saha, {\it {Confronting the Galactic Center
  Gamma Ray Excess with a Light Scalar Dark Matter}},  {\em JCAP} {\bf 1502}
  (2015), no.~02 035, [\href{http://xxx.lanl.gov/abs/1405.0206}{{\tt
  arXiv:1405.0206}}].

\bibitem{Boehm:2014bia}
C.~Boehm, M.~J. Dolan, and C.~McCabe, {\it {A Weighty Interpretation of the
  Galactic Centre Excess}},  {\em Phys.Rev.} {\bf D90} (2014), no.~2 023531,
  [\href{http://xxx.lanl.gov/abs/1404.4977}{{\tt arXiv:1404.4977}}].

\bibitem{Ko:2014gha}
P.~Ko, W.-I. Park, and Y.~Tang, {\it {Higgs Portal Vector Dark Matter for
  $\mathinner{\mathrm{GeV}}$ Scale $\gamma$-ray Excess from Galactic Center}},
  {\em JCAP} {\bf 1409} (2014) 013,
  [\href{http://xxx.lanl.gov/abs/1404.5257}{{\tt arXiv:1404.5257}}].

\bibitem{Abdullah:2014lla}
M.~Abdullah, A.~DiFranzo, A.~Rajaraman, T.~M. Tait, P.~Tanedo, {\em et.~al.},
  {\it {Hidden On-Shell Mediators for the Galactic Center $\gamma$-ray
  Excess}},  {\em Phys.Rev.} {\bf D90} (2014), no.~3 035004,
  [\href{http://xxx.lanl.gov/abs/1404.6528}{{\tt arXiv:1404.6528}}].

\bibitem{Martin:2014sxa}
A.~Martin, J.~Shelton, and J.~Unwin, {\it {Fitting the Galactic Center
  Gamma-Ray Excess with Cascade Annihilations}},  {\em Phys.Rev.} {\bf D90}
  (2014), no.~10 103513, [\href{http://xxx.lanl.gov/abs/1405.0272}{{\tt
  arXiv:1405.0272}}].

\bibitem{Berlin:2014pya}
A.~Berlin, P.~Gratia, D.~Hooper, and S.~D. McDermott, {\it {Hidden Sector Dark
  Matter Models for the Galactic Center Gamma-Ray Excess}},  {\em Phys. Rev.}
  {\bf D90} (2014), no.~1 015032,
  [\href{http://xxx.lanl.gov/abs/1405.5204}{{\tt arXiv:1405.5204}}].

\bibitem{McDermott:2014rqa}
S.~D. McDermott, {\it {Lining Up the Galactic Center Gamma-Ray Excess}},  {\em
  Phys. Dark Univ.} {\bf 7-8} (2014) 12--15,
  [\href{http://xxx.lanl.gov/abs/1406.6408}{{\tt arXiv:1406.6408}}].

\bibitem{Huang:2014cla}
J.~Huang, T.~Liu, L.-T. Wang, and F.~Yu, {\it Supersymmetric subelectroweak
  scale dark matter, the galactic center gamma-ray excess, and exotic decays of
  the 125 gev higgs boson},  {\em Phys. Rev.} {\bf D90} (2014), no.~11 115006,
  [\href{http://xxx.lanl.gov/abs/1407.0038}{{\tt arXiv:1407.0038}}].

\bibitem{Balazs:2014jla}
C.~Bal\'azs and T.~Li, {\it {Simplified Dark Matter Models Confront the Gamma
  Ray Excess}},  {\em Phys. Rev.} {\bf D90} (2014), no.~5 055026,
  [\href{http://xxx.lanl.gov/abs/1407.0174}{{\tt arXiv:1407.0174}}].

\bibitem{Wang:2014elb}
L.~Wang and X.-F. Han, {\it {A Simplified 2Hdm with a Scalar Dark Matter and
  the Galactic Center Gamma-Ray Excess}},  {\em Phys. Lett.} {\bf B739} (2014)
  416--420, [\href{http://xxx.lanl.gov/abs/1406.3598}{{\tt arXiv:1406.3598}}].

\bibitem{Cheung:2014lqa}
C.~Cheung, M.~Papucci, D.~Sanford, N.~R. Shah, and K.~M. Zurek, {\it {NMSSM
  Interpretation of the Galactic Center Excess}},  {\em Phys. Rev.} {\bf D90}
  (2014), no.~7 075011, [\href{http://xxx.lanl.gov/abs/1406.6372}{{\tt
  arXiv:1406.6372}}].

\bibitem{Detmold:2014qqa}
W.~Detmold, M.~McCullough, and A.~Pochinsky, {\it {Dark Nuclei I: Cosmology and
  Indirect Detection}},  {\em Phys. Rev.} {\bf D90} (2014), no.~11 115013,
  [\href{http://xxx.lanl.gov/abs/1406.2276}{{\tt arXiv:1406.2276}}].

\bibitem{Arina:2014yna}
C.~Arina, E.~Del~Nobile, and P.~Panci, {\it {Dark Matter with
  Pseudoscalar-Mediated Interactions Explains the Dama Signal and the Galactic
  Center Excess}},  {\em Phys. Rev. Lett.} {\bf 114} (2015) 011301,
  [\href{http://xxx.lanl.gov/abs/1406.5542}{{\tt arXiv:1406.5542}}].

\bibitem{Han:2014nba}
T.~Han, Z.~Liu, and S.~Su, {\it {Light Neutralino Dark Matter: Direct/Indirect
  Detection and Collider Searches}},  {\em JHEP} {\bf 08} (2014) 093,
  [\href{http://xxx.lanl.gov/abs/1406.1181}{{\tt arXiv:1406.1181}}].

\bibitem{Cline:2014dwa}
J.~M. Cline, G.~Dupuis, Z.~Liu, and W.~Xue, {\it {The Windows for Kinetically
  Mixed Z'-Mediated Dark Matter and the Galactic Center Gamma Ray Excess}},
  {\em JHEP} {\bf 08} (2014) 131,
  [\href{http://xxx.lanl.gov/abs/1405.7691}{{\tt arXiv:1405.7691}}].

\bibitem{Basak:2014sza}
T.~Mondal and T.~Basak, {\it {Class of Higgs-Portal Dark Matter Models in the
  Light of Gamma-Ray Excess from Galactic Center}},  {\em Phys. Lett.} {\bf
  B744} (2015) 208--212, [\href{http://xxx.lanl.gov/abs/1405.4877}{{\tt
  arXiv:1405.4877}}].

\bibitem{Hardy:2014dea}
E.~Hardy, R.~Lasenby, and J.~Unwin, {\it {Annihilation Signals from Asymmetric
  Dark Matter}},  {\em JHEP} {\bf 07} (2014) 049,
  [\href{http://xxx.lanl.gov/abs/1402.4500}{{\tt arXiv:1402.4500}}].

\bibitem{Guo:2014gra}
J.~Guo, J.~Li, T.~Li, and A.~G. Williams, {\it {NMSSM Explanations of the
  Galactic Center Gamma Ray Excess and Promising Lhc Searches}},  {\em Phys.
  Rev.} {\bf D91} (2015), no.~9 095003,
  [\href{http://xxx.lanl.gov/abs/1409.7864}{{\tt arXiv:1409.7864}}].

\bibitem{Yu:2014pra}
J.-H. Yu, {\it {Vector Fermion-Portal Dark Matter: Direct Detection and
  Galactic Center Gamma-Ray Excess}},  {\em Phys. Rev.} {\bf D90} (2014), no.~9
  095010, [\href{http://xxx.lanl.gov/abs/1409.3227}{{\tt arXiv:1409.3227}}].

\bibitem{Cahill-Rowley:2014ora}
M.~Cahill-Rowley, J.~Gainer, J.~Hewett, and T.~Rizzo, {\it {Towards a
  Supersymmetric Description of the Fermi Galactic Center Excess}},  {\em JHEP}
  {\bf 02} (2015) 057, [\href{http://xxx.lanl.gov/abs/1409.1573}{{\tt
  arXiv:1409.1573}}].

\bibitem{Banik:2014eda}
A.~D. Banik and D.~Majumdar, {\it {Low Energy Gamma Ray Excess Confronting a
  Singlet Scalar Extended Inert Doublet Dark Matter Model}},  {\em Phys. Lett.}
  {\bf B743} (2015) 420--427, [\href{http://xxx.lanl.gov/abs/1408.5795}{{\tt
  arXiv:1408.5795}}].

\bibitem{Bell:2014xta}
N.~F. Bell, S.~Horiuchi, and I.~M. Shoemaker, {\it {Annihilating Asymmetric
  Dark Matter}},  {\em Phys. Rev.} {\bf D91} (2015), no.~2 023505,
  [\href{http://xxx.lanl.gov/abs/1408.5142}{{\tt arXiv:1408.5142}}].

\bibitem{Okada:2014usa}
N.~Okada and O.~Seto, {\it {Galactic Center Gamma-Ray Excess from
  Two-Higgs-Doublet-Portal Dark Matter}},  {\em Phys. Rev.} {\bf D90} (2014),
  no.~8 083523, [\href{http://xxx.lanl.gov/abs/1408.2583}{{\tt
  arXiv:1408.2583}}].

\bibitem{Frank:2014bma}
M.~Frank and S.~Mondal, {\it {Light neutralino dark matter in $U(1)'$ models}},
   {\em Phys. Rev.} {\bf D90} (2014), no.~7 075013,
  [\href{http://xxx.lanl.gov/abs/1408.2223}{{\tt arXiv:1408.2223}}].

\bibitem{Hooper:2012cw}
D.~Hooper, N.~Weiner, and W.~Xue, {\it {Dark Forces and Light Dark Matter}},
  {\em Phys. Rev.} {\bf D86} (2012) 056009,
  [\href{http://xxx.lanl.gov/abs/1206.2929}{{\tt arXiv:1206.2929}}].

\bibitem{Carlson:2012qc}
E.~Carlson, D.~Hooper, T.~Linden, and S.~Profumo, {\it {Testing the Dark Matter
  Origin of the Wmap-Planck Haze with Radio Observations of Spiral Galaxies}},
  {\em JCAP} {\bf 1307} (2013) 026,
  [\href{http://xxx.lanl.gov/abs/1212.5747}{{\tt arXiv:1212.5747}}].

\bibitem{Cholis:2012fr}
I.~Cholis, {\it {Searching for the High-Energy Neutrino Counterpart Signals:
  the Case of the Fermi Bubbles Signal and of Dark Matter Annihilation in the
  Inner Galaxy}},  {\em Phys. Rev.} {\bf D88} (2013), no.~6 063524,
  [\href{http://xxx.lanl.gov/abs/1206.1607}{{\tt arXiv:1206.1607}}].

\bibitem{Carlson:2014nra}
E.~Carlson, D.~Hooper, and T.~Linden, {\it {Improving the Sensitivity of
  Gamma-Ray Telescopes to Dark Matter Annihilation in Dwarf Spheroidal
  Galaxies}},  {\em Phys. Rev.} {\bf D91} (2015), no.~6 061302,
  [\href{http://xxx.lanl.gov/abs/1409.1572}{{\tt arXiv:1409.1572}}].

\bibitem{Fields:2014pia}
B.~D. Fields, S.~L. Shapiro, and J.~Shelton, {\it {Galactic Center Gamma-Ray
  Excess from Dark Matter Annihilation: is There a Black Hole Spike?}},  {\em
  Phys. Rev. Lett.} {\bf 113} (2014) 151302,
  [\href{http://xxx.lanl.gov/abs/1406.4856}{{\tt arXiv:1406.4856}}].

\bibitem{Cholis:2014fja}
I.~Cholis, D.~Hooper, and T.~Linden, {\it {A Critical Reevaluation of Radio
  Constraints on Annihilating Dark Matter}},  {\em Phys. Rev.} {\bf D91}
  (2015), no.~8 083507, [\href{http://xxx.lanl.gov/abs/1408.6224}{{\tt
  arXiv:1408.6224}}].

\bibitem{Silverwood:2014yza}
H.~Silverwood, C.~Weniger, P.~Scott, and G.~Bertone, {\it {A Realistic
  Assessment of the Cta Sensitivity to Dark Matter Annihilation}},  {\em JCAP}
  {\bf 1503} (2015), no.~03 055, [\href{http://xxx.lanl.gov/abs/1408.4131}{{\tt
  arXiv:1408.4131}}].

\bibitem{Agrawal:2014oha}
P.~Agrawal, B.~Batell, P.~J. Fox, and R.~Harnik, {\it {WIMPs at the Galactic
  Center}},  {\em JCAP} {\bf 1505} (2015) 011,
  [\href{http://xxx.lanl.gov/abs/1411.2592}{{\tt arXiv:1411.2592}}].

\bibitem{Geringer-Sameth:2014yza}
A.~Geringer-Sameth, S.~M. Koushiappas, and M.~Walker, {\it {Dwarf Galaxy
  Annihilation and Decay Emission Profiles for Dark Matter Experiments}},  {\em
  Astrophys. J.} {\bf 801} (2015), no.~2 74,
  [\href{http://xxx.lanl.gov/abs/1408.0002}{{\tt arXiv:1408.0002}}].

\bibitem{Brooks:2014qya}
A.~Brooks, {\it {Re-Examining Astrophysical Constraints on the Dark Matter
  Model}},  {\em Annalen Phys.} {\bf 526} (2014), no.~7-8 294--308,
  [\href{http://xxx.lanl.gov/abs/1407.7544}{{\tt arXiv:1407.7544}}].

\bibitem{Cholis:2014lta}
I.~Cholis, D.~Hooper, and T.~Linden, {\it {Challenges in Explaining the
  Galactic Center Gamma-Ray Excess with Millisecond Pulsars}},  {\em JCAP} {\bf
  1506} (2015), no.~06 043, [\href{http://xxx.lanl.gov/abs/1407.5625}{{\tt
  arXiv:1407.5625}}].

\bibitem{Cholis:2014noa}
I.~Cholis, D.~Hooper, and T.~Linden, {\it {A New Determination of the Spectra
  and Luminosity Function of Gamma-Ray Millisecond Pulsars}},
  \href{http://xxx.lanl.gov/abs/1407.5583}{{\tt arXiv:1407.5583}}.

\bibitem{Cirelli:2014lwa}
M.~Cirelli, D.~Gaggero, G.~Giesen, M.~Taoso, and A.~Urbano, {\it {Antiproton
  constraints on the GeV gamma-ray excess: a comprehensive analysis}},  {\em
  JCAP} {\bf 1412} (2014), no.~12 045,
  [\href{http://xxx.lanl.gov/abs/1407.2173}{{\tt arXiv:1407.2173}}].

\bibitem{Zhou:2014lva}
B.~Zhou, Y.-F. Liang, X.~Huang, X.~Li, Y.-Z. Fan, {\em et.~al.}, {\it {GeV
  Excess in the Milky Way: the Role of Diffuse Galactic Gamma Ray Emission
  Template}},  \href{http://xxx.lanl.gov/abs/1406.6948}{{\tt arXiv:1406.6948}}.

\bibitem{Bringmann:2014lpa}
T.~Bringmann, M.~Vollmann, and C.~Weniger, {\it {Updated Cosmic-Ray and Radio
  Constraints on Light Dark Matter: Implications for the GeV Gamma-Ray Excess
  at the Galactic Center}},  \href{http://xxx.lanl.gov/abs/1406.6027}{{\tt
  arXiv:1406.6027}}.

\bibitem{Portillo:2014ena}
S.~K.~N. Portillo and D.~P. Finkbeiner, {\it {Sharper Fermi Lat Images:
  Instrument Response Functions for an Improved Event Selection}},  {\em
  Astrophys. J.} {\bf 796} (2014), no.~1 54,
  [\href{http://xxx.lanl.gov/abs/1406.0507}{{\tt arXiv:1406.0507}}].

\bibitem{Petrovic:2014uda}
J.~Petrovic, P.~D. Serpico, and G.~Zaharijas, {\it {Galactic Center Gamma-Ray
  "Excess" from an Active Past of the Galactic Centre?}},  {\em JCAP} {\bf
  1410} (2014), no.~10 052, [\href{http://xxx.lanl.gov/abs/1405.7928}{{\tt
  arXiv:1405.7928}}].

\bibitem{Cholis:2015dea}
I.~Cholis, C.~Evoli, F.~Calore, T.~Linden, C.~Weniger, and D.~Hooper, {\it {The
  Galactic Center GeV Excess from a Series of Leptonic Cosmic-Ray Outbursts}},
  \href{http://xxx.lanl.gov/abs/1506.0511}{{\tt arXiv:1506.0511}}.

\bibitem{Carlson:2014cwa}
E.~Carlson and S.~Profumo, {\it {Cosmic Ray Protons in the Inner Galaxy and the
  Galactic Center Gamma-Ray Excess}},  {\em Phys. Rev.} {\bf D90} (2014), no.~2
  023015, [\href{http://xxx.lanl.gov/abs/1405.7685}{{\tt arXiv:1405.7685}}].

\bibitem{Yoast-Hull:2014cra}
T.~M. Yoast-Hull, J.~S. Gallagher, and E.~G. Zweibel, {\it {The Cosmic Ray
  Population of the Galactic Central Molecular Zone}},  {\em Astrophys. J.}
  {\bf 790} (2014) 86, [\href{http://xxx.lanl.gov/abs/1405.7059}{{\tt
  arXiv:1405.7059}}].

\bibitem{Bernal:2014mmt}
N.~Bernal, J.~E. Forero-Romero, R.~Garani, and S.~Palomares-Ruiz, {\it
  {Systematic uncertainties from halo asphericity in dark matter searches}},
  {\em JCAP} {\bf 1409} (2014) 004,
  [\href{http://xxx.lanl.gov/abs/1405.6240}{{\tt arXiv:1405.6240}}].

\bibitem{Bramante:2014zca}
J.~Bramante and T.~Linden, {\it {Detecting Dark Matter with Imploding Pulsars
  in the Galactic Center}},  {\em Phys. Rev. Lett.} {\bf 113} (2014), no.~19
  191301, [\href{http://xxx.lanl.gov/abs/1405.1031}{{\tt arXiv:1405.1031}}].

\bibitem{Drlica-Wagner:2014yca}
A.~Drlica-Wagner, G.~A. G\'omez-Vargas, J.~W. Hewitt, T.~Linden, and
  L.~Tibaldo, {\it {Searching for Dark Matter Annihilation in the Smith
  High-Velocity Cloud}},  {\em Astrophys. J.} {\bf 790} (2014) 24,
  [\href{http://xxx.lanl.gov/abs/1405.1030}{{\tt arXiv:1405.1030}}].

\bibitem{Cholis:2013ena}
I.~Cholis, D.~Hooper, and S.~D. McDermott, {\it {Dissecting the Gamma-Ray
  Background in Search of Dark Matter}},  {\em JCAP} {\bf 1402} (2014) 014,
  [\href{http://xxx.lanl.gov/abs/1312.0608}{{\tt arXiv:1312.0608}}].

\bibitem{Hooper:2014ysa}
D.~Hooper, T.~Linden, and P.~Mertsch, {\it {What Does the PAMELA Antiproton
  Spectrum Tell Us About Dark Matter?}},  {\em JCAP} {\bf 1503} (2015), no.~03
  021, [\href{http://xxx.lanl.gov/abs/1410.1527}{{\tt arXiv:1410.1527}}].

\bibitem{Geringer-Sameth:2014qqa}
A.~Geringer-Sameth, S.~M. Koushiappas, and M.~G. Walker, {\it {Comprehensive
  Search for Dark Matter Annihilation in Dwarf Galaxies}},  {\em Phys. Rev.}
  {\bf D91} (2015), no.~8 083535,
  [\href{http://xxx.lanl.gov/abs/1410.2242}{{\tt arXiv:1410.2242}}].

\bibitem{Tavakoli:2013zva}
M.~Tavakoli, I.~Cholis, C.~Evoli, and P.~Ullio, {\it {Constraints on Dark
  Matter Annihilations from Diffuse Gamma-Ray Emission in the Galaxy}},  {\em
  JCAP} {\bf 1401} (2014) 017, [\href{http://xxx.lanl.gov/abs/1308.4135}{{\tt
  arXiv:1308.4135}}].

\bibitem{Gaggero:2015nsa}
D.~Gaggero, M.~Taoso, A.~Urbano, M.~Valli, and P.~Ullio, {\it {Towards a
  Realistic Astrophysical Interpretation of the Galactic Center Excess}},
  \href{http://xxx.lanl.gov/abs/1507.0612}{{\tt arXiv:1507.0612}}.

\bibitem{Lee:2014mza}
S.~K. Lee, M.~Lisanti, and B.~R. Safdi, {\it {Distinguishing Dark Matter from
  Unresolved Point Sources in the Inner Galaxy with Photon Statistics}},  {\em
  JCAP} {\bf 1505} (2015), no.~05 056,
  [\href{http://xxx.lanl.gov/abs/1412.6099}{{\tt arXiv:1412.6099}}].

\bibitem{Lee:2015fea}
S.~K. Lee, M.~Lisanti, B.~R. Safdi, T.~R. Slatyer, and W.~Xue, {\it {Evidence
  for Unresolved Gamma-Ray Point Sources in the Inner Galaxy}},
  \href{http://xxx.lanl.gov/abs/1506.0512}{{\tt arXiv:1506.0512}}.

\bibitem{Bartels:2015aea}
R.~Bartels, S.~Krishnamurthy, and C.~Weniger, {\it {Strong Support for the
  Millisecond Pulsar Origin of the Galactic Center GeV Excess}},
  \href{http://xxx.lanl.gov/abs/1506.0510}{{\tt arXiv:1506.0510}}.

\bibitem{Brandt:2015ula}
T.~D. Brandt and B.~Kocsis, {\it {Disrupted Globular Clusters Can Explain the
  Galactic Center Gamma Ray Excess}},  {\em Astrophys. J.} {\bf 812} (2015),
  no.~1 15, [\href{http://xxx.lanl.gov/abs/1507.0561}{{\tt arXiv:1507.0561}}].

\bibitem{Carlson:2015ona}
E.~Carlson, T.~Linden, and S.~Profumo, {\it {Putting Things Back Where They
  Belong: Tracing Cosmic-Ray Injection with H2}},
  \href{http://xxx.lanl.gov/abs/1510.0469}{{\tt arXiv:1510.0469}}.

\bibitem{Linden:2015qha}
T.~Linden, {\it {Known Radio Pulsars Do Not Contribute to the Galactic Center
  Gamma-Ray Excess}},  \href{http://xxx.lanl.gov/abs/1509.0292}{{\tt
  arXiv:1509.0292}}.

\bibitem{Dutta:2015ysa}
B.~Dutta, Y.~Gao, T.~Ghosh, and L.~E. Strigari, {\it {Confronting Galactic
  Center and Dwarf Spheroidal Gamma-Ray Observations with Cascade Annihilation
  Models}},  {\em Phys. Rev.} {\bf D92} (2015), no.~7 075019,
  [\href{http://xxx.lanl.gov/abs/1508.0598}{{\tt arXiv:1508.0598}}].

\bibitem{deBoer:2015kta}
W.~de~Boer, I.~Gebauer, S.~Kunz, and A.~Neumann, {\it {Evidence for a Hadronic
  Origin of the Fermi Bubbles and the Galactic Excess}},
  \href{http://xxx.lanl.gov/abs/1509.0531}{{\tt arXiv:1509.0531}}.

\bibitem{Petrovic:2014xra}
J.~Petrovi{\'c}, P.~D. Serpico, and G.~Zaharijas, {\it {Millisecond pulsars and
  the Galactic Center gamma-ray excess: the importance of luminosity function
  and secondary emission}},  {\em JCAP} {\bf 1502} (2015), no.~02 023,
  [\href{http://xxx.lanl.gov/abs/1411.2980}{{\tt arXiv:1411.2980}}].

\bibitem{blobs}
{\url{http://fermi.gsfc.nasa.gov/ssc/data/access/lat/Model\_details/Pass7\_galactic.html}}.

\bibitem{scipy}
E.~Jones, T.~Oliphant, and P.~Peterson, {\it et. al., scipy: Open source
  scientific tools for python},  2001.

\bibitem{ipython}
F.~P\'erez and B.~E. Granger, {\it Ipython: a system for interactive scientific
  computing},  {\em Comput. Sci. Eng.} {\bf 9} (2007), no.~3 21--29.

\bibitem{Gorski:2004by}
K.~M. Gorski, E.~Hivon, A.~J. Banday, B.~D. Wandelt, F.~K. Hansen, M.~Reinecke,
  and M.~Bartelman, {\it {Healpix - a Framework for High Resolution
  Discretization, and Fast Analysis of Data Distributed on the Sphere}},  {\em
  Astrophys. J.} {\bf 622} (2005) 759--771,
  [\href{http://xxx.lanl.gov/abs/astro-ph/0409513}{{\tt astro-ph/0409513}}].

\end{thebibliography}\endgroup

\end{document}